\documentclass[12pt, a4paper]{article}
\usepackage[height=25cm,width=16cm]{geometry}
\usepackage{feynmp}
\usepackage{amsmath}
\usepackage{ascmac}
\usepackage{dcolumn}
\usepackage{bm,here}
\usepackage{subfig}
\usepackage{comment}
\usepackage{ifpdf}
\usepackage{slashed}
\usepackage{colortbl}
\usepackage{color}
\usepackage{ulem}
\usepackage{comment}
\usepackage{braket}
\usepackage[mathscr]{eucal}
\usepackage[sort&compress, numbers, merge]{natbib}
\usepackage{cancel}

\ifpdf
\usepackage{graphicx}     %   use package without driver option
\usepackage[bookmarksopen,colorlinks=true,linkcolor=bblue,citecolor=ppink,urlcolor=ppink]{hyperref}
\else     % For (p)LaTeX + dvipdfmx
\usepackage[dvipdfmx]{graphicx}     %   use package with driver option
\usepackage[dvipdfmx,bookmarksopen,colorlinks=true,linkcolor=bblue,citecolor=ppink,urlcolor=ppink]{hyperref}
\fi

\usepackage{multicol}
\definecolor{red}{rgb}{1,0,0}
\definecolor{ppink}{rgb}{0.921545,0.440586,0.687243}
\definecolor{bblue}{rgb}{0.400000,0.400000,1.000000}
\usepackage[charter]{mathdesign}
\usepackage{soul}
\usepackage{wrapfig}
\usepackage{arydshln}

\newcommand\blfootnote[1]{%
	\begingroup
	\renewcommand\thefootnote{}\footnote{#1}%
	\addtocounter{footnote}{-1}%
	\endgroup
}

\begin{document}

%%%%%%%%%%%%%%%%%%%%%%%%%%%%%
%%%%%%%%%%% Title %%%%%%%%%%%
%%%%%%%%%%%%%%%%%%%%%%%%%%%%%
\begin{titlepage}

\begin{flushright}
%    \hfill IPMU21-xxx \\
\end{flushright}
%\vskip 1cm ~

\begin{center}

	\vskip 1.5cm
	{\large \bf Self-Interacting Sub-GeV Dark Matter with Strong MeV Gamma-ray}

	\vskip 2.0cm
	{\large Yu Watanabe \blfootnote{yuuwatanabe.phys@gmail.com} }
	
	\vskip 2.0cm
	$^1${\sl Department of Physics and Astronomy, University of California, Los Angeles, California, 90095-1547, USA} \\ [.3em]

    \vskip 3.5cm
    \begin{abstract}
        \noindent
        Sub-GeV dark matter (DM) with $s$-channel resonant self-scattering provides a promising framework for addressing small-scale structure problems. However, models that also account for the observed relic abundance through the same resonance are strongly constrained by current $\gamma$-ray observations, since the associated signals are significantly enhanced. To overcome this limitation, we propose a framework in which the relic abundance and self-scattering are governed independently by two distinct mediators. As a concrete realization, we present a singlet scalar DM model in which self-scattering is mediated by a vector boson associated with a gauged baryon number, while the relic density is determined by forbidden annihilation into dark Higgs bosons that generate the gauge boson mass. By imposing cosmological, experimental, and theoretical constraints, We identify viable parameter regions that reproduce the observed relic density, alleviate small-scale problems, and remain consistent with current bounds. Notably, the model predicts multiple distinctive MeV $\gamma$-ray signals, a significant fraction of which will be testable with next-generation MeV $\gamma$-ray telescopes, including the Compton Spectrometer and Imager (COSI).
%Emphasizing minimal framework
    \end{abstract}
		
\end{center}
		
\end{titlepage}

%%%%%%%%%%%%%%%%%%%%%%%%%%%%%%%%
%%%%%%%%%%% Contents %%%%%%%%%%%
%%%%%%%%%%%%%%%%%%%%%%%%%%%%%%%%
\tableofcontents
\newpage
\setcounter{page}{1}

%%%%%%%%%%%%%%%%%%%%%%%%%%%%%%%%%%%%
%%%%%%%%%%% Introduction %%%%%%%%%%%
%%%%%%%%%%%%%%%%%%%%%%%%%%%%%%%%%%%%
\section{Introduction} 
\label{sec: intro} 

Revealing the nature of dark matter (DM) is a central open problem in astrophysics, cosmology, and particle physics. A well-motivated candidate is thermal DM, which was once in equilibrium with Standard Model (SM) particles in the early universe and whose relic abundance is set by freeze-out\,\cite{Srednicki:1988ce, Bernstein:1985th}. Electroweak-scale thermal DM, or weakly interacting massive particles (WIMPs), have been the subject of extensive searches, yet the absence of conclusive signals has increasingly constrained their parameter space\,\cite{Arcadi:2017kky, Giusti:1998gz}. This motivates the exploration of alternative mass regimes, particularly MeV-scale thermal DM, often referred to as sub-GeV DM. Sub-GeV DM typically produces MeV $\gamma$-rays, where experimental sensitivity is limited—the so-called ``MeV gap.'' Recent advances, however, have led to new mission proposals to fill this gap, most notably NASA’s Compton Spectrometer and Imager (COSI) satellite\,\cite{Tomsick:2021wed, Tomsick:2023aue}, scheduled for launch in 2027. It is thus timely to study the phenomenology of sub-GeV DM and its implications for future $\gamma$-ray observations.

Sub-GeV DM is subject to stringent cosmological constraints\,\cite{Slatyer:2015jla, Kawasaki:2021etm}. Evading these constraints typically requires velocity-dependent annihilation processes\,\cite{Watanabe:2025pvc}, such as $p$-wave\,\cite{Pospelov:2007mp, Matsumoto:2018acr}, forbidden\,\cite{DAgnolo:2015ujb, DAgnolo:2020mpt}, or $s$-channel resonance-enhanced annihilation\,\cite{Feng:2017drg, Bernreuther:2020koj, Binder:2022pmf, Brahma:2023psr}. Among these mechanisms, resonant annihilation is particularly appealing for addressing small-scale structure problems. Inconsistencies in DM density profiles at galactic centers (GCs), such as the core–cusp and diversity problems, can be alleviated if DM exhibits a sufficiently large self-scattering cross section, which thermalizes itself in GCs\,\cite{Spergel:1999mh, Kamada:2016euw}. Resonant annihilation often enhances self-scattering as well, yielding the desired velocity dependence of the cross section\,\cite{Chu:2018fzy}.

However, models that attempt to explain both the observed DM abundance and small-scale structure problems via the same resonance are strongly constrained by current telescopes\,\cite{Duch:2017nbe, Binder:2022pmf}. Explaining the observed relic density requires a sizable annihilation rate, while resolving small-scale issues demands a resonance at DM relative velocities of $\sim 10^{-3}$. Both effects significantly enhance DM-induced $\gamma$-ray signals from the GC. In this article, we propose a new framework to address these challenges. One of the simplest sub-GeV DM scenarios is the singlet DM extension of the SM with a new gauged U(1) symmetry\,\cite{Watanabe:2025pvc}. The new gauge boson mixes with SM neutral bosons and mediates interactions between DM and SM particles. The minimal UV completion of this model often introduces a complex scalar, the dark Higgs, which generates the gauge boson mass through its vacuum expectation value (VEV). While the dark Higgs is usually assumed to decouple from DM phenomenology, it can be as light as the DM itself and significantly enrich the dynamics\,\cite{Darme:2017glc, Wojcik:2021xki, Dutra:2025cwn}, mixing with the SM Higgs and acting as a scalar mediator. We propose treating relic abundance and self-scattering requirements independently through these two mediators, thereby decoupling the resonance needed for self-scattering from the large annihilation cross section required by freeze-out. This separation allows the annihilation cross section to evade current telescope bounds while still benefiting from resonant enhancement, yielding strong $\gamma$-ray signals potentially observable by near-future experiments. As an illustrative example, we focus on a model in which singlet scalar DM self-scatters via the resonance of a vector mediator associated with the gauged baryon number U(1)$_{\rm B}$ and undergoes forbidden annihilation into dark Higgs bosons. The proposed framework can be straightforwardly extended to alternative U(1) charge assignments, different spin combinations of DM and mediators, and other annihilation mechanisms.
%A minimal model for WIMP+MED
%Why U(1)B?

In this article, we construct the general UV-complete Lagrangian of the model. Sub-GeV DM is subject to multiple constraints: cosmological (from Big Bang Nucleosynthesis (BBN) and Cosmic Microwave Background (CMB) observations), experimental (from accelerators as well as direct and indirect DM searches), and theoretical (from perturbative unitarity and vacuum stability). By scanning over all free parameters in the Lagrangian, we find viable parameter regions that simultaneously reproduce today’s DM relic density via the freeze-out mechanism, address small-scale structure problems, and satisfy all constraints. We further find that, when projected onto observables relevant for upcoming experiments, a significant fraction of these parameter sets can be probed by COSI. The model predicts four types of MeV $\gamma$-ray signals: DM annihilation via the vector mediator resonance into electron-positron pairs produces a continuum signal through final state radiation (FSR), while annihilation into neutral pions and photons produces line signals. In addition, DM forbiddenly annihilates into the scalar mediator, which subsequently decays into electron-positron pairs and two photons. Each of these channels features a distinctive spectrum and sizable flux, making the model an attractive target for the rapidly advancing field of MeV $\gamma$-ray observations.

This paper is organized as follows. In Section\,\ref{sec: scenario}, we present the setup of the model, namely the general UV-complete Lagrangian of the singlet scalar DM extension of the gauged U(1)$_{\rm B}$ SM with a dark Higgs, detailing the effective interactions at the MeV scale and providing formulae for various phenomenologically relevant quantities. In Section\,\ref{sec: conditions and constraints}, we discuss the relevant conditions and constraints on the model. Section\,\ref{sec: The Status of Self-Interacting Sub-GeV Dark Matter} implements these conditions and constraints, identifies viable parameter regions that satisfy all requirements, and evaluates their detection prospects in upcoming experiments. Finally, Section\,\ref{sec: Summary} summarizes our findings and conclusions.

%%%%%%%%%%%%%%%%%%%%%%%%%%%%%%%%%%
%%%%%%%%%%% model %%%%%%%%%%%
%%%%%%%%%%%%%%%%%%%%%%%%%%%%%%%%%%
\section{Self-interacting sub-GeV dark matter}
\label{sec: scenario}

We discuss a self-interacting sub-GeV DM, particularly focusing on a singlet scalar DM realized in the gauged U(1)$_{\rm B}$ extension of the SM, where B denotes the baryon number. The general Lagrangian describing the DM and its interactions with SM particles is given by:
\begin{align}
    &\mathscr{L} =
      \mathcal{L}_{\rm SM}
    + \mathcal{L}_{\rm B}
    + \mathcal{L}_{\rm DM},
    \nonumber \\
    &\mathcal{L}_{\rm B} =
    -\frac{1}{4} (V_{\mu \nu}')^2
    +|D_\mu S|^2
    +\mathcal{L}_{\rm f}\,(\psi, \eta, \chi)
    -\frac{\xi}{2} V_{\mu \nu}'\,B^{\mu \nu}
    -g_{\rm B}\,V_\mu'\,\sum_f q_f \bar{f} \gamma^\mu f
    \nonumber \\
    & \qquad~~
    - \frac{\lambda_{S}}{2}(|S|^2 - v_S^2/2)^2
    - \lambda_{HS}(|H|^2 - v_H^2/2)\,(|S|^2 - v_S^2/2), 
    \nonumber \\
    &\mathcal{L}_{\rm DM} =
    |D_\mu \varphi|^2 - m_\varphi^2 |\varphi|^2
    - \frac{\lambda_\varphi}{4} |\varphi|^4
    \nonumber \\
    & \qquad~~
    -\lambda_{H \varphi} (|H|^2 - v_H^2/2)\, |\varphi|^2
    -\lambda_{S \varphi} (|S|^2 - v_S^2/2) \,|\varphi|^2,
    \label{eq: lagrangian 1}
\end{align}
where $\mathcal{L}_{\rm SM}$ denotes SM Lagrangian; $V_{\mu\nu}'$ is the field strength tensor of the U(1)$_{\rm B}$ gauge boson $V_\mu'$; $S$ is the dark "Higgs" field that spontaneously breaks the U(1)$_{\rm B}$ symmetry; $\psi$, $\eta$ and $\chi$ represents fermions introduced to ensure the model is anomaly-free\,\cite{Duerr:2013lka, Duerr:2013dza, Duerr:2014wra}; $f$ is SM fermions; $H$ is the SM Higgs doublet; and $\varphi$ is the scalar DM. The quantum numbers of the new particles are summarized in Table\,\ref{tab: representations}. A $Z_2$ symmetry is imposed to guarantee DM stability, under which all other particles are even. The covariant derivative acting on a particle $f$ is given by  
\begin{equation}
    D_\mu = \partial_\mu + i g_s T^a_G G_\mu^a + i g T^a_W W_\mu^a + i g' Y_f B_\mu + i g_{\rm B} q_f V_\mu',
\end{equation}
where $G_\mu^a$, $W_\mu^a$, and $B_\mu$ are the gauge fields of SU(3)$_c$, SU(2)$_L$, and U(1)$_Y$ groups in the SM. The generators of the SU(3)$_c$ and SU(2)$_L$ groups are denoted by $T^a_G$ and $T^a_W$, while $Y_f$ and $q_f$ represent the hypercharge and the baryon number of the particle $f$. The corresponding gauge couplings of the gauge fields are denoted by $g_s$, $g$, $g'$, and $g_{\rm B}$, respectively.

After the electroweak (EW) symmetry breaking, we take the unitary gauge with $H = (0, v_H + h')^T/\sqrt{2}$, where $v_H \simeq 246$\,GeV is its VEV. The U(1)$_{\rm B}$ symmetry is broken by the scalar $S = (v_S + \varsigma')/\sqrt{2}$, with $v_S$ being its VEV. This breaking generates masses for the new particles: the U(1)$_{\rm B}$ gauge boson acquires $m_{V'} \sim g_{\rm B}\,v_S$ and the dark higgs boson obtains $m_{\varsigma'} \sim \sqrt{\lambda_S}\,v_S$. In this article, we focus on the case where the DM, the U(1)$_{\rm B}$ gauge boson and dark higgs boson all have masses of $\mathcal{O}(1$--$100)\,\mathrm{MeV}$. As discussed in Sec.~\ref{sec: Constraints from Accelerators}, constraints from accelerator experiments require $g_{\rm B} \lesssim 10^{-3.5} (m_{V'}/\mathrm{GeV})$, which in turn implies $v_S \gtrsim \mathrm{few~TeV}$. The term $\mathcal{L}_{\rm f}$ encodes interactions involving the anomaly-canceling fermions; we omit its explicit form as it is irrelevant for our phenomenological discussion. These fermions acquire masses through Yukawa couplings with $S$, and thanks to the large value of $v_S$, they can be sufficiently heavy to evade current collider constraints.\footnote{
    The $Z_2$ symmetry, i.e., $\psi_{L/R} \to -\psi_{L/R}$, $\eta_{L/R} \to -\eta_{L/R}$, and $\chi_{L/R} \to -\chi_{L/R}$ remains unbroken even after the U(1)$_{\rm B}$ and EW symmetry breaking. Thus, the lightest of these fermions may also contribute to the present dark matter abundance. This contribution can be suppressed if the reheating temperature of the universe is sufficiently low.
}

At the MeV scale, the dark higgs boson mixes with the SM Higgs, thereby acting as a scalar mediator. Similarly, the U(1)$_{\rm B}$ gauge boson mixes with the neutral SM gauge bosons $W^3_\mu$ and $B_\mu$, serving as a vector mediator. In the following subsections, we present the effective interactions of these mediators at the MeV scale and provide expressions for physical quantities that are directly relevant to our phenomenological analysis.

\begin{table}[t]
    \centering
    \begin{tabular}{c|ccccc}
        & SU(3)$_c$ & SU(2)$_L$ & U(1)$_Y$ & U(1)$_{\rm B}$ & $Z_2$ \\
        \hline
        $S$ & {\bf 1} & {\bf 1} & 0 & $-3$ & + \\
        $\psi_{L/R}$ & {\bf 1} & {\bf 2} & $-1/2$ & $B_1/B_2$ & $+$ \\
        $\eta_{R/L}$ & {\bf 1} & {\bf 1} & $-1$ & $B_1/B_2$ & $+$ \\
        $\chi_{R/L}$ & {\bf 1} & {\bf 1} & $0$ & $B_1/B_2$ & $+$ \\
        $\varphi$ & {\bf 1} & {\bf 1} & 0 & $q_\varphi$ & $-$ \\
        \hline
    \end{tabular}
    \caption{\small\sl Quantum numbers of the new particles. $B_{1}$ is arbitrary with $B_2 - B_1 = 3$}
    \label{tab: representations}
\end{table}

%%%%%%%%%%%%%%%%%%%%%%%%%%%
%%%%%%%%%%%%%%%%%%%%%%%%%%%
\subsection{Effective interactions including the scalar mediator}
\label{subsec: Effective interactions scalar}

After the EW and U(1)$_{\rm B}$ symmetries breaking, the quadratic terms of $h'$ and $\varsigma'$ are obtained as
\begin{align}
    {\cal L}_S \, \supset \,
	\frac{1}{2}
	\begin{pmatrix} h' & \varsigma' \end{pmatrix}
	\begin{pmatrix} m^2_{h' h'} & m^2_{h' \varsigma'} \\ m^2_{h' \varsigma'} & m^2_{\varsigma' \varsigma'} \end{pmatrix}
	\begin{pmatrix} h' \\ \varsigma' \end{pmatrix}
	\, = \,
	\frac{1}{2}
	\begin{pmatrix} h & \varsigma \end{pmatrix}
	\begin{pmatrix} m_h^2 & 0 \\ 0 & m_\varsigma^2 \end{pmatrix}
	\begin{pmatrix} h \\ \varsigma \end{pmatrix},
\end{align}
where $m_{h' h'}^2 = \lambda_H v_H^2$, $m_{h' \varsigma'}^2 = \lambda_{HS} v_H v_S$ and $m_{\varsigma' \varsigma'}^2 = \lambda_{S} v_S^2$, respectively. The coupling constant $\lambda_H$ is from the Higgs potential in the SM; ${\cal L}_{\rm SM} \supset -(\lambda_H/2)(|H|^2 - v_H^2/2)^2$. The mixing matrix diagonalizing the mass matrix and relating the states $(h', \varsigma')$ to $(h, \varsigma)$ is written as
\begin{align}
	\begin{pmatrix} h \\ \varsigma \end{pmatrix} =
	\begin{pmatrix} \cos\theta & -\sin\theta \\ \sin\theta & \cos\theta \end{pmatrix}
	\begin{pmatrix} h' \\ \varsigma' \end{pmatrix}.
\end{align}
Mass eigenstates and the mixing angle are $m_h^2,\,m_\varsigma^2 = [(m_{h' h'}^2 + m_{\varsigma' \varsigma'}^2) \pm \{(m_{h' h'}^2 - m_{\varsigma' \varsigma'}^2)^2 + 4m_{h' \varsigma'}^4\}^{1/2}]/2$ and $\tan\, (2\theta) = 2m_{h' \varsigma'}^2/(m_{\varsigma' \varsigma'}^2 - m_{h' h'}^2)$. Since the mixing angle $\theta$ is constrained to be small, as will be discussed in the following sections, $h$ is almost the SM Higgs boson, whose mass is given by $m_h \simeq 125$\,GeV as confirmed by the collider (LHC) experiment. Then the Lagrangian in eq.~\eqref{eq: lagrangian 1} contains the following interactions:
\begin{align}
    {\cal L}_S \, \supset
	&
    -\frac{s_\theta \varsigma + c_\theta h}{v_H} \sum_f m_f \bar{f}f
	+\left[\frac{s_\theta \varsigma + c_\theta h}{v_H} + \frac{(s_\theta \varsigma + c_\theta h)^2}{2v_H^2}\right] \left(2m_W^2 W_\mu^\dagger W^\mu + m_Z^2 Z_\mu Z^\mu\right)
    \nonumber \\
    &
    -\frac{C_{h h h}}{3!} h^3
	-\frac{C_{\varsigma h h}}{2} \varsigma h^2
	-\frac{C_{\varsigma \varsigma h}}{2} \varsigma^2 h
	-\frac{C_{\varsigma \varsigma \varsigma }}{3!} \varsigma^3
    \nonumber\\
	&
	-\frac{C_{h h h h}}{4!} h^4
	-\frac{C_{\varsigma h h h}}{3!} \varsigma h^3
        -\frac{C_{\varsigma \varsigma h h}}{4} \varsigma^2 h^2
        -\frac{C_{\varsigma \varsigma \varsigma h}}{3!} \varsigma^3 h
        -\frac{C_{\varsigma \varsigma \varsigma \varsigma}}{4!} \varsigma^4
    \\
	&
    -C_{h \varphi \varphi}\, h |\varphi|^2
    -C_{\varsigma \varphi \varphi} \, \varsigma |\varphi|^2
    -\frac{C_{h h \varphi \varphi}}{2} h^2 |\varphi|^2
    -C_{h \varsigma \varphi \varphi} \, h \varsigma |\varphi|^2
    -\frac{C_{\varsigma \varsigma \varphi \varphi}}{2} \varsigma^2 |\varphi|^2
    -\frac{\lambda_\varphi}{4} |\varphi|^4,
    \nonumber
	\label{eq: S interactions}
\end{align}
where $c_\theta \equiv \cos\theta$ and $s_\theta \equiv \sin\theta$, and $m_f$ denotes the mass of field $f$. Scalar interaction coefficients—those involving $h$, $\varsigma$, and $\varphi$—are given in Appendix~\ref{app: scalar interactions}. $W_\mu$ and $Z_\mu$ denote the SM weak gauge bosons.\footnote{The interactions involving the $Z$ boson are affected by mixing with the vector mediator, as discussed in Subsec.~\ref{subsec: Effective interactions vector}. However, these effects are highly suppressed by the small mixing parameter $\xi \ll 1$ and are therefore irrelevant for the subsequent discussion.}

%%%%%%%%%%%%%%%%%%%%%%%%%%%
%%%%%%%%%%%%%%%%%%%%%%%%%%%
\subsection{Effective interactions including the vector mediator}
\label{subsec: Effective interactions vector}

After the EW and U(1)$_{\rm B}$ symmetries breaking, the quadratic terms of the neutral gauge bosons are obtained as
\begin{align}
    &{\cal L}_V \supset
    \frac{1}{2} (W^3_\mu,\,B_\mu,\,V'_\mu)
    \left[ (\Box g^{\mu \nu} - \partial^\mu \partial^\nu)\,{\cal K} + {\cal M}\,g^{\mu \nu} \right]
    (W^3_\nu,\,B_\nu,\,V'_\nu)^T,
    \nonumber \\
    & {\cal K} \equiv
    \begin{pmatrix} 1 & 0 & 0 \\ 0 & 1 & \xi \\ 0 & \xi & 1 \end{pmatrix},
    \qquad
    {\cal M} \equiv
    \begin{pmatrix} g^2 v_H^2/4 & -g g' v_H^2/4 & 0 \\ -g g' v_H^2/4 &  g^{\prime 2} v_H^2/4 & 0 \\ 0 & 0 & 9 g_{\rm B}^2 v_S^2 \end{pmatrix}.
\end{align}
The redefinition of the gauge fields, as well as the diagonalization of the above mass matrix ${\cal M}$, give mass eigenstates having canonical kinetic terms,
\begin{align}
    &
    X {\cal K} X^T = {\bf 1},
    \qquad
    X {\cal M} X^T = {\rm diag}(m_Z^2,\,0,\,m_V^2),
    \qquad
    (Z_\mu,\,A_\mu,\,V_\mu)^T = (X^{-1})^T (W^3_\mu,\,B_\mu,\,V'_\mu)^T,
    \nonumber \\
    & \qquad\qquad\qquad\quad
    X \simeq
    \begin{pmatrix}
        c_W & - s_W & \displaystyle \xi \frac{s_W m_Z^2}{m_Z^2 - m_V^2} \\
        s_W & c_W & 0 \\
        \displaystyle -\xi \frac{c_W \, s_W m_Z^2}{m_Z^2 - m_V^2} & \displaystyle \xi \frac{m_V^2 - c^2_W m_Z^2}{m_Z^2 - m_V^2} & 1
    \end{pmatrix},
\end{align}
under the assumption that $\xi \ll 1$, where $c_W \equiv \cos\theta_W$ and $s_W \equiv \sin\theta_W$ with $\theta_W$ being the Weinberg angle. Here, $A$ denotes the photon and $m_V$ is the mass of the vector mediator particle $V$. This mixing shifts the coupling of vector mediator to U(1)$_{\rm B}$ charged particle as:
\begin{align}
    g_{\rm B} q_f \rightarrow g_{\rm B} q_f
    -g' \xi\,Q_{f} \frac{c_W^2 m_Z^2 - m_V^2}{m_Z^2 - m_V^2}
    -g' \xi\,T_f\frac{m_V^2}{m_Z^2 - m_V^2},
\end{align}
Here, $Q_f$ is the electric charge and $T_f$ is the third component of the weak isospin. At energies well below the GeV scale, the light quarks should be expressed in terms of hadronic degrees of freedom, whose interactions can be derived using chiral perturbation theory\,\cite{Scherer:2002tk, Coogan:2021sjs}. Consequently, the effective interactions including the vector mediator at the MeV scale can be written as:
\begin{align}
    {\cal L}_V \, \supset
	&
    -g_{Ve} \, V^\mu \, i\,(\pi^+ \overleftrightarrow{\partial_\mu} \pi^-)
    + g_{Ve}^2 V^\mu V_\mu \pi^+\pi^-
    + 2 e g_{Ve} \, A^\mu V_\mu \pi^+\pi^-
    - g_{Ve} \,(\bar{e} \slashed{V} e +   \bar{\mu} \slashed{V} \mu)
    \nonumber \\
    &
    -\frac{\epsilon^{\mu\nu\rho\sigma}}{8\pi^2 f_\pi} (\partial_\mu \pi^0) \, \left[ e(g_{\rm B} - g_{Ve}) \left\{ (\partial_\nu \, A_\rho) \, V_\sigma + (\partial_\nu \, V_\rho) \, A_\sigma \right\} + g_{Ve} (2 g_{\rm B} - g_{Ve}) \, (\partial_\nu \, V_\rho) \, V_\sigma\right]
    \nonumber \\
    &
    + m_V^2 V^\mu V_\mu \, \left[\frac{c_\theta \varsigma - s_\theta h}{v_S} + \frac{(c_\theta \varsigma - s_\theta h)^2}{2v_S^2}\right] 
    - g_\varphi \, V^\mu \, i\,(\varphi^* \overleftrightarrow{\partial_\mu} \varphi) 
    + g_\varphi^2 \, V^\mu V_\mu |\varphi|^2, 
	\label{eq: V interactions}
\end{align}
where $g_{Ve} \equiv \xi c_W e$, where $e$ is the QED coupling, and $g_\varphi \equiv g_{\rm B} \, q_\varphi$. $\pi^\pm$ and $\pi^0$ denote the charged and neutral pion fields, respectively, and $f_\pi$ is the pion decay constant. The fields $e$ and $\mu$ represent the electron and muon, respectively.

%%%%%%%%%%%%%%%%%%%%%%%%%%%
%%%%%%%%%%%%%%%%%%%%%%%%%%%
\subsection{Phenomenological quantities}
\label{subsec: Phenomenological quantities}

We concentrate on the parameter space where DM annihilates forbiddenly into the scalar mediator, while annihilations into SM particles and DM self-interactions are resonantly enhanced by the vector mediator. These processes occur under the conditions $m_\varphi \leq m_\varsigma \equiv m_\varphi (1 + v_{\rm th}^2/8)$ and $2m_\varphi \leq m_V \equiv 2m_\varphi (1 + v_{\rm R}^2/8)$,  where $v_{\rm th}$ ($v_{\rm R}$) denotes the relative velocity of the incident DM at the threshold (resonance with the vector mediator propagator). To keep the discussion simple i.e. avoid complications from DM annihilations into multiple hadrons, we restrict the DM mass to be below $100\,{\rm MeV}$.  
In this subsection we summarize the expressions for decay widths, annihilation cross sections, and scattering processes relevant for our phenomenological analysis.

%%%%%%%%%%%%%%%%%%%%%%%%%%%
%%%%%%%%%%%%%%%%%%%%%%%%%%%
\pmb{Decay of the scalar mediator:}  
The scalar mediator behaves like a light SM Higgs boson: Its decay width into SM particles is given by $\Gamma\,(\varsigma \to {\rm SMs}) = s_\theta^2 \, \Gamma(h_{\rm SM} \to {\rm SMs})$, with $m_h$ replaced by $m_\varsigma$. For $m_\varsigma < 2m_e$, the dominant channel is decay into two photons, with partial width $\Gamma(\varsigma \to \gamma \gamma)$ computed using Ref.~\cite{Leutwyler:1989tn}. In the range $2 m_e \leq m_\varsigma < 2 m_\mu$, the mediator predominantly decays into an electron–positron pair, with leading-order width
\begin{equation}
	\Gamma(\varsigma \to e^- e^+) = s_\theta^2 \frac{m_e^2 m_\varsigma}{8 \pi v_H^2} 
	\left( 1 - \frac{4 m_e^2}{m_\varsigma^2} \right)^{3/2}.
	\label{eq: electrons}
\end{equation}
Radiative decays of the type $\varsigma \to e^- e^+ \gamma$ also occur. The differential partial width is\,\cite{Coogan:2019qpu}
{\small
\begin{align}
    &\frac{d\Gamma(\varsigma \to e^- e^+ \gamma)}{dE_\gamma}
    =\frac{2 \alpha}{\pi m_\varsigma} \Gamma(\varsigma \to e^- e^+)
    \times {\rm FSRS}(2E_\gamma/m_\varsigma, m_e/m_\varsigma),
    \label{eq: s to eegamma}
    \\
    &{\rm FSRS}(x, \mu)
    =\frac{2(1-x-6\mu^2)+(x+4\mu^2)^2}{x\,(1 - 4\mu^2)^{3/2}}
    \log\left[\frac{1 + v_\mu(x)}{1 - v_\mu(x)}\right]
    -\frac{2(1 - 4\mu^2)(1 - x)}{x\,(1 - 4\mu^2)^{3/2}}v_\mu(x),
    \nonumber
\end{align}
}\noindent
where $\alpha$ is the fine-structure constant of the electromagnetic interaction, $E_\gamma$ is the photon energy, and $v_\mu(x) \equiv \sqrt{1 - 4 \mu^2/(1 - x)}$. In the mass range considered here, the scalar mediator does not decay into DM or the vector mediator.

%%%%%%%%%%%%%%%%%%%%%%%%%%%
%%%%%%%%%%%%%%%%%%%%%%%%%%%
\pmb{Decay of the vector mediator:}  
For $m_{V} \leq 2m_e$, the vector mediator primarily decays into three photons, with approximate width\,\cite{Pospelov:2008jk, McDermott:2017qcg, Linden:2024fby}
\begin{align}
    \Gamma(V \to \gamma \gamma \gamma) \simeq
    \frac{17\alpha^3 g_{Ve}^2}{2^{9} 3^6 5^3 \pi^4} \frac{m_{V}^9}{m_e^8},
\end{align}
valid for $m_{V} \ll 2 m_e$.\footnote{For our analysis, we use the exact one-loop result of Ref.~\cite{McDermott:2017qcg}.} Above threshold ($m_{V} \geq 2m_e$), the mediator decays to $e^- e^+$ and $e^- e^+ \gamma$. The leading-order partial width into electrons is
{\small
\begin{align}
    &\Gamma(V \to e^- e^+) = \frac{m_{V}}{12 \pi}
        g_{Ve}^2 \left(1 + \frac{2 m_e^2}{m_{V}^2}\right)
        \left(1 - \frac{4 m_e^2}{m_{V}^2}\right)^{1/2},
    \label{eq: V-SM coupling}
\end{align}
}\noindent
and the differential radiative width is\,\cite{Coogan:2019qpu}
{\small
\begin{align}
    &
    \frac{d\Gamma(V \to e^- e^+ \gamma)}{dE_\gamma} =
    \frac{2 \alpha}{\pi m_{V}} \Gamma(V \to e^- e^+)
    \times {\rm FSRV}(2E_\gamma/m_{V}, m_e/m_{V}),
    \label{eq: V to eegamma}
    \\
    &
    {\rm FSRV}(x, \mu) =
    \frac{1 + (1 - x)^2 - 4\mu^2 (x + 2 \mu^2)}{x\,\sqrt{1 - 4\mu^2}\,(1+2\mu^2)}
    \log\left[\frac{1 + v_\mu(x)}{1 - v_\mu(x)}\right]
    -\frac{1 + (1 - x)^2 + 4\mu^2 (1 - x)}{x\,\sqrt{1 - 4\mu^2}\,(1 + 2\mu^2)}
    v_\mu(x),
    \nonumber
\end{align}
}\noindent
For $m_{V} \geq m_{\pi^0}$, the vector mediator also decays into $\pi^0 \gamma$\,\cite{Coogan:2019qpu, Coogan:2021sjs}, with partial width
\begin{align}
    \Gamma(V \to \pi^0 \gamma) =
    \frac{\alpha\,(g_{\rm B} - g_{Ve})^2 (m_{V}^2 - m_{\pi^0}^2)^3}{1536\,\pi^4 f_\pi^2 m_{V}^3}.
\label{eq: decay pi0gamma}
\end{align}
In addition, the mediator can decay invisibly into DM pairs:
\begin{align}
    \Gamma(V \to \varphi \varphi^*) =
    \frac{g_\varphi^2 m_{V}}{384\pi} v_{\rm R}^3.
\end{align}

%%%%%%%%%%%%%%%%%%%%%%%%%%%
%%%%%%%%%%%%%%%%%%%%%%%%%%%
\pmb{Decay of the Higgs boson:}
Higgs couplings to SM particles are modified by mixing with the dark higgs boson. Its decay width into SM states is therefore $\Gamma\,(h \to {\rm SMs}) = c_\theta^2\,\Gamma(h_{\rm SM} \to {\rm SMs})$. In addition, the Higgs can decay into DM and scalar mediators:
\begin{align}
	\Gamma(h \to \varphi \varphi^*) &= \frac{C_{h \varphi \varphi}^2}{32 \pi m_h} \left(1 - \frac{4 m_\varphi^2}{m_h^2} \right)^{1/2}, \nonumber \\
	\Gamma(h \to \varsigma \varsigma) &= \frac{C_{ \varsigma \varsigma h}^2}{32 \pi m_h} \left(1 - \frac{4 m_\varsigma^2}{m_h^2} \right)^{1/2}.
	\label{eq: hinv}
\end{align}
Since $s_\theta$ is tightly constrained (Sec.~\ref{sec: Constraints from Accelerators}), decays into vector mediators are negligible.

%%%%%%%%%%%%%%%%%%%%%%%%%%%
%%%%%%%%%%%%%%%%%%%%%%%%%%%
\pmb{DM annihilation:}
The annihilation cross section into an SM final state $f_{\rm SM}$ is
\begin{align}
    \sigma v\,(\varphi \varphi^* \to f_{\rm SM}) &\simeq \frac{96 \pi}{\sqrt{1-4m_\varphi^2/s}}
    \frac{[\Gamma\,(V \to \varphi \varphi^*)
    \, \Gamma\,(V \to f_{\rm SM})]_{m_V^2 \to s}}{(s - m_V^2)^2 + s\,\Gamma_{V,\text{tot}}^2(s)}\\
    &\simeq \frac{192 \pi}{v_\varphi m_\varphi^4}
    \frac{[\Gamma\,(V \to \varphi \varphi^*)
    \, \Gamma\,(V \to f_{\rm SM})]_{m_V^2 \to s}}{(v_\varphi^2 - v_{\rm R}^2)^2 + 16\Gamma_{V,\text{tot}}^2(s)/m_V^2},
    \label{eq: Ann Xsection}
\end{align}
where $s$ is the center-of-mass energy, $v_\varphi \equiv 2\sqrt{1 - 4 m_\varphi^2/s}$ is the relative velocity of the incident DM, and $\Gamma_{V,\text{tot}}$ is the total width of $V$.\footnote{The $s$-channel diagrams involving exchange of the scalar mediator or Higgs boson contribute at $s$-wave to DM annihilation into SM final states. Although potentially relevant away from the resonance ($v_\varphi < v_{\rm R}$), they are suppressed by the small $s_\theta$ and the electron Yukawa coupling, and are therefore neglected, as they provide a negligible contribution to the constraints discussed in the following sections.} DM can also annihilate into scalar mediators:
{\small
\begin{align}
    \frac{d\sigma v\,(\varphi \varphi^* \to \varsigma \varsigma)}{d\Omega}
    = \frac{\sqrt{1 - 4 m_\varsigma^2/s}}{32 \pi^2\,s}
    \left|
        C_{\varsigma \varsigma \varphi \varphi}
        +
        \frac{C_{\varsigma \varphi \varphi}\,C_{\varsigma \varsigma \varsigma}}{s -m_\varsigma^2}
        +
        \frac{C_{h \varphi \varphi}\,C_{\varsigma \varsigma h}}{s - m_h^2}
        +
        \frac{C_{\varsigma \varphi \varphi}^2}{t-m_\varphi^2}
        +
        \frac{C_{\varsigma \varphi \varphi}^2}{u-m_\varphi^2}
    \right|^2,
    \label{eq: Ann forbidden}
\end{align}
}\noindent
with $t$ and $u$ the Mandelstam variables.  

\begin{figure}[t]
    \centering
    \includegraphics[keepaspectratio, scale=0.65]{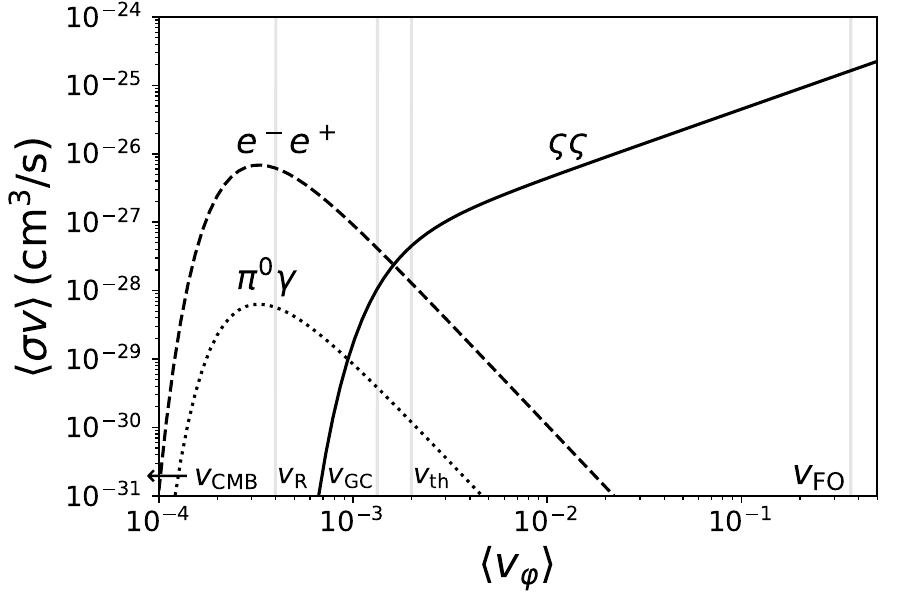}
    \caption{\small \sl The velocity dependence of DM annihilation cross sections into various final states is shown for the parameter set $(m_\varphi, v_{\rm R}, v_{\rm th}, \xi, g_{\rm B}, C_{\varsigma\varphi\varphi}) = (69~\mathrm{MeV}, 4\times 10^{-4}, 2\times 10^{-3}, 10^{-11}, 2\times 10^{-8})$ as an illustrative example, with other parameters having no impact on this behavior. Here, $v_{\rm FO} \sim \mathcal{O}(0.1)$ denotes the DM relative velocity at freeze-out, and $v_{\rm GC} \sim 400$\,km/s corresponds to the DM velocity at the GC~\cite{Lacroix:2020lhn}. The DM velocity at the recombination era, $v_{\rm CMB}$, is extremely small and lies outside the range shown in the figure.}
    \label{fig: sv}
\end{figure}

The above annihilation cross sections, averaged over the relative-velocity distribution $f(v_\varphi, v_0) \equiv 4 v_\varphi^2 e^{-v_\varphi^2/v_0^2}/(\sqrt{\pi}\,v_0^3)$ with mean velocity $\langle v_\varphi \rangle = 2v_0/\sqrt{\pi}$, are summarized in Fig.\,\ref{fig: sv}. For $v_\varphi \geq v_{\rm th}$, annihilation into scalar mediators proceeds efficiently, reaching a sizable value at freeze-out. For $v_\varphi < v_{\rm th}$, the cross section is highly suppressed due to insufficient center-of-mass energy. Nevertheless, the resulting $\varsigma$ subsequently decays into $\gamma\gamma$ or $e^- e^+ \gamma$, producing substantial line or continuum $\gamma$-ray signals in the present universe. For $v_\varphi \gtrsim v_{\rm R}$, annihilation into SM final states is resonantly enhanced, yielding prominent line or continuum $\gamma$-rays via the $\pi^0 \gamma$ or $e^- e^+ \gamma$ channels. Away from resonance ($v_\varphi < v_{\rm R}$), the cross section is exponentially suppressed, as it becomes unlikely to hit the propagator pole.

%%%%%%%%%%%%%%%%%%%%%%%%%%%
%%%%%%%%%%%%%%%%%%%%%%%%%%%
\pmb{DM self-scattering:}
The self-scattering cross sections are
\begin{align}
    \sigma \, (\varphi \varphi^* \to \varphi \varphi^*) &\simeq 2\sigma_0 + \frac{192 \pi}{v_\varphi^2 m_\varphi^4}
    \frac{[\Gamma\,(V \to \varphi \varphi^*)]^2_{m_V^2 \to s}}{(v_\varphi^2 - v_{\rm R}^2)^2 + 16\Gamma_{V,\text{tot}}^2(s)/m_V^2}\\
    \sigma \, (\varphi \varphi \to \varphi \varphi) &= 
    \sigma \, (\varphi^* \varphi^* \to \varphi^* \varphi^*) 
    \simeq \sigma_0,
    \label{eq: cs SI}
\end{align}
with $\sigma_0 \equiv \lambda_\varphi^2/(128\pi m_\varphi^2)$.\footnote{
Velocity-independent contributions from the $s$-, $t$-, and $u$-channel diagrams are neglected, as the corresponding couplings are suppressed relative to $\lambda_\varphi$, as discussed in the following sections. Moreover, the contribution from interference between the contact and $s$-channel diagrams is also negligible~\cite{Chu:2018fzy}.} The cross section consists of a velocity-independent offset $\sigma_0$ and a resonantly enhanced component 
centered at $v_{\rm R}$.

%%%%%%%%%%%%%%%%%%%%%%%%%%%
%%%%%%%%%%%%%%%%%%%%%%%%%%%
\pmb{DM scattering with an electron:}
The cross section for DM scattering off electrons in the zero-momentum transfer limit is 
\begin{align}
    & \sigma_e(\varphi e \to \varphi e) \simeq
    \frac{m_e^2}{4 \pi (m_\varphi + m_e)^2}
    \left[ \frac{m_e}{v_H}\left(
    s_\theta \frac{C_{\varsigma \varphi \varphi}}{m_\varsigma^2}
    +c_\theta \frac{C_{h \varphi \varphi}}{m_h^2}
    \right)
    - 2 g_{Ve} g_\varphi \frac{m_\varphi}{m_V^2}
    \right]^2.
    \label{eq: DM-e scattering}
\end{align}

%%%%%%%%%%%%%%%%%%%%%%%%%%%%%%%%%%%%%%%%%%%%%%%%
%%%%%%%%%%% Conditions & Constraints %%%%%%%%%%%
%%%%%%%%%%%%%%%%%%%%%%%%%%%%%%%%%%%%%%%%%%%%%%%%
\section{Conditions and constraints}
\label{sec: conditions and constraints}

We investigate the phenomenology of self-interacting sub-GeV DM introduced in Section~\ref{sec: scenario}. We begin by examining the role of DM self-interactions in addressing the small-scale crisis in cosmic structure formation. Next, we analyze the conditions under which the model reproduces the observed relic abundance via the freeze-out mechanism. For these conditions, we introduce likelihood functions to identify the favored parameter regions. We then consider constraints on these parameters from cosmological observations, such as those related to the CMB and BBN, as well as bounds from experimental searches for new particles in accelerator experiments, underground detectors, and astrophysical observations. Additional constraints, including theoretical considerations and limits from supernova cooling, are also discussed. A more detailed discussion from a broader perspective can be found in \cite{Watanabe:2025pvc}.

%%%%%%%%%%%%%%%%%%%%%%%%%%%
%%%%%%%%%%%%%%%%%%%%%%%%%%%
\subsection{Self-scattering condition}
\label{subsec: self-scattering}

\begin{figure}[t]
    \centering    \includegraphics[keepaspectratio, scale=0.65]{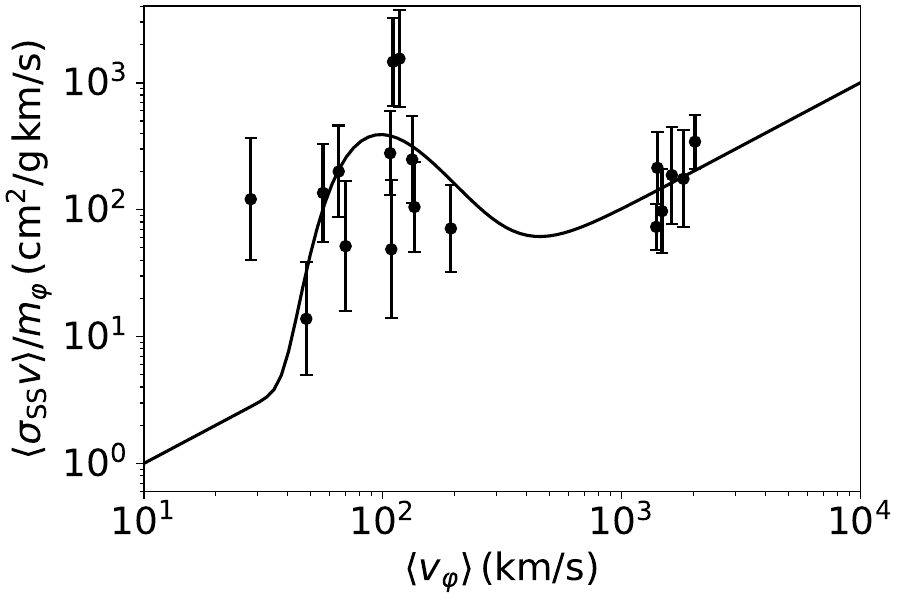}
    \caption{\small \sl The velocity dependence of the DM self-scattering cross section for the parameter set $(m_\varphi, v_R, g_\varphi, \sigma_0/m_\varphi) = (100~\mathrm{MeV}, 4\times 10^{-4}, 4 \times 10^{-3}, 0.067~\mathrm{cm}^2/\mathrm{g})$, with other parameters having no impact on this behavior, is shown as a black solid line. The cross sections required to alleviate small-scale structure issues, inferred from kinematical observations, are indicated by black dots with error bars~\cite{Kaplinghat:2015aga}.}
    \label{fig: self-scattering}
\end{figure}

The core--cusp and diversity problems can be alleviated if DM exhibits a sufficiently large self-scattering cross section\,\cite{Spergel:1999mh, Tulin:2017ara}. The required magnitude and velocity dependence of the cross section have been quantitatively analyzed in Ref.\,\cite{Kaplinghat:2015aga}, as illustrated in Fig.\,\ref{fig: self-scattering}. The data points are taken from five galaxy clusters\,\cite{Newman:2012nw}, seven low-surface-brightness spiral galaxies\,\cite{KuziodeNaray:2007qi}, and six dwarf galaxies from the THINGS sample\,\cite{Oh:2010ea}. 

In the scenario considered here, where the present-day DM consists equally of $\varphi$ and $\varphi^*$, the effective self-scattering cross section is given by
\begin{align}
    \sigma_{\rm SS} =
    \frac{1}{4} \sigma (\varphi \varphi \to \varphi \varphi)
    +\frac{1}{2} \sigma (\varphi \varphi^* \to \varphi \varphi^*)
    +\frac{1}{4} \sigma (\varphi^* \varphi^* \to \varphi^* \varphi^*).
    \label{eq: self-scattering cross-section}
\end{align}
I require the model parameters to yield a self-scattering cross section consistent with the data, quantified by the likelihood
\begin{eqnarray}
    -2\ln {\mathcal L}_{\rm SS} = \sum_i
    \left\{
        \frac{(\text{Center value})_i - \log_{10}[\langle \sigma_{\rm SS} v \rangle_{i}/m_\varphi]}{(\text{Error bar})_i}
    \right\}^2,
\label{eq: likelihood self-scattering}
\end{eqnarray}
where ${\cal L}_{\rm SS}$ is the likelihood function for the self-scattering condition, and the index $i$ labels each galaxy or cluster in the data, with its corresponding central value, uncertainty, and mean DM velocity as shown in Fig.\,\ref{fig: self-scattering}.

For DM masses at the MeV scale, the self-scattering condition favors specific model parameters, assuming that the vector mediator width is dominated by its decay into a DM pair\,\cite{Chu:2018fzy}.\footnote{In the viable parameter region, $V$ decays primarily into DM, as discussed in the following section.} In particular, a narrow resonance is required, with its total width scaling as $\Gamma_{V,\text{tot}} \propto m_\varphi^4$. Furthermore, the parameters $v_{\rm R}$ and $\sigma_0/m_\varphi$ are preferred to be approximately $100\,\mathrm{km/s}$ and $0.1\,\mathrm{cm}^2/\mathrm{g}$, respectively, as indicated by the solid line in Fig.\,\ref{fig: self-scattering}.

%%%%%%%%%%%%%%%%%%%%%%%%%%%
%%%%%%%%%%%%%%%%%%%%%%%%%%%
\subsection{Relic abundance condition}
\label{subsec: Relic abundance}

The evolution of the DM phase-space density $f_{\varphi}$ in the early universe is governed by the Boltzmann equation, $L[f_{\varphi}] = C[f_{\varphi}]$, where $L$ and $C$ denote the Liouville operator and collision terms, respectively\,\cite{Kolb:1990vq, Gondolo:1990dk}. Solving this equation numerically is generally time-consuming. An alternative is to consider the first few moments of $f_{\varphi}$, i.e.\ the so-called hydrodynamical approach\,\cite{Binder:2017rgn, Binder:2021bmg}. The zeroth moment corresponds to the DM number density, $n_{\varphi}(t)$. Assuming that DM remains in thermal equilibrium with the SM bath during freeze-out closes the Boltzmann hierarchy at this level, reducing the equation to the familiar form
\begin{align}
    \dot{n}_{\varphi^{(*)}} + 3 H n_{\varphi^{(*)}} =
    -\langle \sigma v \rangle \,\left[n_{\varphi^{(*)}}^2 - ( n_{\varphi^{(*)}}^{\rm eq})^2 \right],
    \label{eq: nBE}
\end{align}
where $H$ is the Hubble parameter and $n_{\varphi}^{\rm eq} \sim (m_{\varphi} T)^{3/2}\exp (-m_{\varphi}/T)$ is the equilibrium DM number density, with $T$ denoting the temperature of the universe. In our model, the annihilation cross section $\langle \sigma v \rangle$ receives two contributions (see Sec.\,\ref{subsec: Phenomenological quantities}): annihilation into SM particles via the vector mediator resonance, and forbidden annihilation into scalar mediators. As discussed in Sec.\,\ref{subsec: self-scattering}, $v_{\rm R} \sim 100\,\mathrm{km/s} \sim v_{\rm GC}$ is required to address small-scale structure problems, for which the resonant contribution is strongly constrained by indirect detection (see Sec.\,\ref{sec: Constraints from Indirect Detection}). Our numerical analysis confirms that its effect on the relic density is negligible. Consequently, the relic abundance is determined almost entirely by the forbidden annihilation channel. For simplicity, we impose thermal equilibrium between the dark sector and the SM bath during freeze-out\footnote{The vector mediator does not play a role in maintaining thermal equilibrium with the SM bath, as its number density is Boltzmann suppressed due to its relatively large mass, and its couplings to SM states are small.}:
\begin{align}
    \langle \sigma v (\varphi^{(*)} \, {\rm SM} \to \varphi^{(*)} \, {\rm SM}) \rangle_{{\rm FO}}\,n_{\rm SM}^{\rm eq}(T_{\rm FO}) > H(T_{\rm FO}) 
    \quad {\rm or}\quad
    \langle \Gamma (\varsigma \to {\rm SM}) \rangle_{{\rm FO}}
    > H(T_{\rm FO}), 
    \label{eq: kinetic equilibrium}
\end{align}
where $\Gamma$ denotes the decay width including the Lorentz factor, and $\langle ... \rangle_{\rm FO}$ indicates the thermal average evaluated at the freeze-out temperature $T_{\rm FO}$. The freeze-out temperature $T_{\rm FO}$ is defined through the condition $n_{\varphi}(T_{\rm FO}) = 2.5\,n_{\varphi}^{\rm eq}(T_{\rm FO})$. DM tends to remain in thermal equilibrium with the scalar mediator due to crossing symmetry, but thermally decouples around freeze-out because of the nearly degenerate mass spectrum. This effect induces an approximately $10\%$ uncertainty in the predicted relic density when $v_{\rm th}\ll 1$~\cite{Aboubrahim:2023yag}. For each parameter point, we first check whether the kinetic equilibrium condition \eqref{eq: kinetic equilibrium} is satisfied; points that fail are discarded. Otherwise, the DM relic abundance $\Omega_{\rm TH}\,h^2$ is computed by numerically solving eq.\,\eqref{eq: nBE}.

The CMB precisely measures the DM relic density as $\Omega_{\rm DM} h^2 \simeq 0.12$ with ${\cal O}(1\%)$ uncertainty~\cite{Planck:2018vyg}. On the theory side, several additional sources of uncertainty affect the calculation of the DM abundance. The relativistic degrees of freedom in the SM plasma contribute an ${\cal O}(0.1\%)$ effect~\cite{Saikawa:2020swg}. For MeV-scale DM, freeze-out occurs around neutrino decoupling, which modifies the temperature evolution of SM species. When DM couples predominantly to electrons and photons—as in our setup—this effect can induce up to a $10\%$ shift in the predicted relic density~\cite{Li:2023puz}. We therefore conservatively adopt a $10\%$ theoretical uncertainty and model the likelihood associated with the relic abundance as
\begin{align}
    -2\ln {\mathcal L}_{\rm RA} =
    \left(
        \frac{\Omega_{\rm DM}\,h^2 - \Omega_{\rm TH}\,h^2}{0.1\,\Omega_{\rm DM}\,h^2}
    \right)^2,
    \label{eq: likelihood relic abundance}
\end{align}
The relic abundance condition requires $\langle \sigma v\rangle_{{\rm FO}} \sim 10^{-26\text{--}25}\,{\rm cm^3/s}$, which fixes correlations among the scalar couplings appearing in eq.\,\eqref{eq: Ann forbidden}.

%%%%%%%%%%%%%%%%%%%%%%%%%%%
%%%%%%%%%%%%%%%%%%%%%%%%%%%
\subsection{Cosmological constraints}
\label{subsec: Cosmological Constraints}

Sub-GeV DM is subject to constraints that differ from those of traditional EW-scale WIMPs, as it is expected to remain in thermal contact with SM particles for an extended period and to continue annihilating, thereby injecting energy into the plasma at late times. These processes can leave observable imprints on cosmological probes, most notably the CMB and BBN. In this subsection, we discuss the resulting constraints.

%%%%%%%%%%%%%%%%%%%%%%%%%%%
%%%%%%%%%%%%%%%%%%%%%%%%%%%
\subsubsection{CMB observation}
\label{subsubsec: Constraints from CMB Observation}

During the recombination epoch, sub-GeV DM continues to inject electromagnetically interacting particles into the SM plasma. This alters the thermal history of recombination by increasing the residual ionization fraction, thereby modifying the CMB anisotropy\,\cite{Slatyer:2015jla, Kawasaki:2021etm, Lopez-Honorez:2013cua}. Since precise CMB observations favor the standard anisotropy without additional energy injection, the DM annihilation cross section at recombination, $\langle \sigma v \rangle_{\rm CMB}$, is constrained as
\begin{align}
\label{eq: CMB limit on annihilation}
    f_{\rm eff\,}(m_{\varphi})\,
    \frac{\langle \sigma v \rangle_{\rm CMB}}{m_{\varphi}} \leq
    4.1 \times 10^{-28}\,[{\rm cm}^3\,{\rm s}^{-1}\,{\rm GeV}^{-1}],
\end{align}
at 95\% C.L. by the Planck collaboration\,\cite{Planck:2018vyg}. Here $f_{\rm eff}(m_\varphi)$ denotes the efficiency with which the annihilation energy is deposited into the plasma, for which we adopt the values from Ref.\,\cite{Slatyer:2015jla}. The DM velocity during recombination can be estimated as $\langle v_\varphi \rangle_{\rm CMB} \simeq 2 \times 10^{-5}\,( T_\gamma/1\,{\rm eV})\,(1\,{\rm MeV}/m_\varphi)\,(10^{-2}/x_{\rm kd})^{1/2}$ with $T_\gamma = 0.235$\,eV and $x_{\rm kd} \equiv T_{\rm kd}/m_\varphi$ the ratio of the kinetic decoupling temperature to the DM mass\,\cite{Essig:2013goa}. The kinetic decoupling temperature is determined by eq.\,(\ref{eq: kinetic equilibrium}). Equation~\eqref{eq: CMB limit on annihilation} implies that for $m_\varphi \simeq 1$--$100$\,MeV, the annihilation cross section must satisfy $\langle \sigma v \rangle_{\rm CMB} \lesssim 10^{-30}$--$10^{-28}$\,cm$^3$/s, far below the canonical thermal relic value $\langle \sigma v \rangle_{{\rm FO}} \sim 10^{-26}$\,cm$^3$/s. However, since $\langle v_\varphi \rangle_{\rm CMB} \ll v_{\rm R}$ and also lies below $v_{\rm th}$ unless the latter is unnaturally small ($v_{\rm th} \lesssim 10^{-5}$), the annihilation cross section at recombination is strongly suppressed, while remaining unsuppressed at freeze-out, as illustrated in Fig.\,\ref{fig: sv}. Consequently, our model can reproduce the observed DM relic abundance while safely evading the stringent CMB constraints.
%I -> we, my -> our?

The existence of new MeV-scale particles --- DM and scalar/vector mediators in our model --- may also affect the physics of neutrino decoupling. If these new particles are sufficiently light to become non-relativistic after the neutrino decoupling epoch, their entropy is transferred exclusively to the electromagnetic plasma. As a result, the photon temperature increases relative to the neutrino temperature, and the expansion rate of the Universe decreases\,\cite{Dolgov:2002wy, Ibe:2018juk}. Precise CMB observations support the standard expansion history without such entropy injection, thereby imposing lower bounds on the masses of new particles\,\cite{Giovanetti:2021izc, Sabti:2021reh, Chu:2022xuh}. This constraint is usually expressed in terms of the effective number of relativistic degrees of freedom,
\begin{align}
    N_{\rm eff} =
    3 \left[ \frac{11}{4}\left( \frac{T_\nu}{T_\gamma}\right)^3 \right]^{4/3},
\end{align}
where $T_\gamma$ and $T_\nu$ denote the photon and neutrino temperatures at recombination. The Planck collaboration reports $N_{\rm eff} = 2.99 \pm 0.17$ at the 1$\sigma$ confidence level\,\cite{Planck:2018vyg}, in excellent agreement with the SM prediction\,\cite{deSalas:2016ztq, Bennett:2020zkv}. The contribution of new particles to $N_{\rm eff}$ can be estimated as
\begin{align}
    N_{\rm eff} \simeq 3 \left\{ 1 + \frac{45}{11\pi^2 T_D^3} \sum_i s_i (T_D)   \right\}^{-4/3},
    \qquad
    s_i(T_D) = h_i(T_D) \frac{2\pi^2}{45} T_D^3,
    \label{eq: electrophilic Neff}
\end{align}
with $h_i(T_D) = (15 x_i^4)/(4 \pi^4) \int^{\infty}_1 dy \,(4y^2 -1)\sqrt{y^2 -1}/(e^{x_i y} - 1)$, where $x_i \equiv m_i/T_D$, $T_D \simeq 1.7\,{\rm MeV}$ is the neutrino decoupling temperature, $g_i$ denotes the number of internal degrees of freedom, and the sum runs over all new particles labeled by $i$\,\cite{Matsumoto:2018acr, Ibe:2018juk}.\footnote{Here we assume that the effect of the new particles enters only through entropy injection, together with instantaneous neutrino decoupling via weak interactions. This approximation yields results in good agreement with more sophisticated treatments using coupled Boltzmann equations for the electromagnetic, neutrino, and dark sectors\,\cite{Escudero:2018mvt, Sabti:2019mhn}.} Comparing eq.\,(\ref{eq: electrophilic Neff}) with the observed $N_{\rm eff}$ yields a lower limit on the DM mass, $m_\varphi \gtrsim 7.6$\,MeV within the parameter space considered in this work.

%%%%%%%%%%%%%%%%%%%%%%%%%%%
%%%%%%%%%%%%%%%%%%%%%%%%%%%
\subsubsection{BBN observation}
\label{subsubsec: Constraints from BBN Observation}

If new particles inject high-energy, electromagnetically interacting particles into the cosmic plasma, they initiate electromagnetic showers that produce energetic photons, which can destroy light elements via photo-disintegration processes~\cite{Kawasaki:1994af, Ellis:1990nb}. Consequently, the DM annihilation cross section is constrained to avoid spoiling the success of standard BBN, $\langle \sigma v \rangle \lesssim 10^{-25}\,{\rm cm^3/s}$ at 95\% C.L.~\cite{Depta:2019lbe, Braat:2024khe}. Although this limit is below the canonical thermal relic cross section required for the observed DM abundance, it may still affect the resonant component of the annihilation cross section if the resonance is enhanced at the photo-disintegration epoch. However, indirect detection constraints (see subsubsection~\ref{sec: Constraints from Indirect Detection}) are even more stringent for the resonant component, since addressing the small-scale crisis requires $v_{\rm R} \sim v_{\rm GC}$, as discussed in subsection~\ref{subsec: self-scattering}. Therefore, we do not further consider this bound.

New light particles can also modify the expansion rate of the Universe and the evolution of photon and neutrino temperatures before their energy densities become Boltzmann suppressed, thereby altering the primordial light-element abundances. However, BBN observations are consistent with the standard thermal history, setting lower bounds on the masses of new particles~\cite{Boehm:2013jpa}. Recent studies report limits of order $\mathcal{O}(0.1)\,{\rm MeV}$~\cite{Boehm:2013jpa, Sabti:2019mhn, Watanabe:2025pvc}, which are weaker than those derived from CMB observations (see section~\ref{subsubsec: Constraints from CMB Observation}). For this reason, we do not consider them further.
\subsection{Experimental constraints}
\label{subsec: Experimental Constraints}

The overall strategy to search for sub-GeV DM is conceptually similar to that for EW-scale WIMPs: signals are sought in underground laboratories (direct detection), at accelerators and colliders, and through astrophysical observations (indirect detection). However, the experimental techniques differ significantly due to the lower DM mass scale and the presence of a light mediator. Below, we summarize the current status and future prospects for probing sub-GeV DM.

%%%%%%%%%%%%%%%%%%%%%%%%%%%
%%%%%%%%%%%%%%%%%%%%%%%%%%%
\subsubsection{Direct detection experiments}
\label{sec: Constraints from Direct Detection}

Direct detection experiments aim to measure recoil signals induced by DM scattering off ordinary matter, such as nuclei or electrons, in deep underground detectors. Conventional EW-scale WIMP searches, which typically use noble-liquid targets such as xenon, lose sensitivity for light DM because the recoil energy is strongly suppressed and often falls below detector thresholds. Alternative strategies are therefore required, including the development of detectors with substantially lower thresholds\,\cite{Kurinsky:2019pgb, Griffin:2020lgd} and the exploitation of the Migdal effect\,\cite{Ibe:2017yqa, Dolan:2017xbu}. Among these, the most sensitive approach is to search for DM scattering off bound electrons\,\cite{Essig:2011nj, Essig:2015cda}, where nearly the entire kinetic energy of the DM can be transferred to the recoiling electron. In what follows, we focus on constraints from DM–electron scattering.

\begin{figure}[t]
    \centering
    \includegraphics[keepaspectratio, scale=0.65]{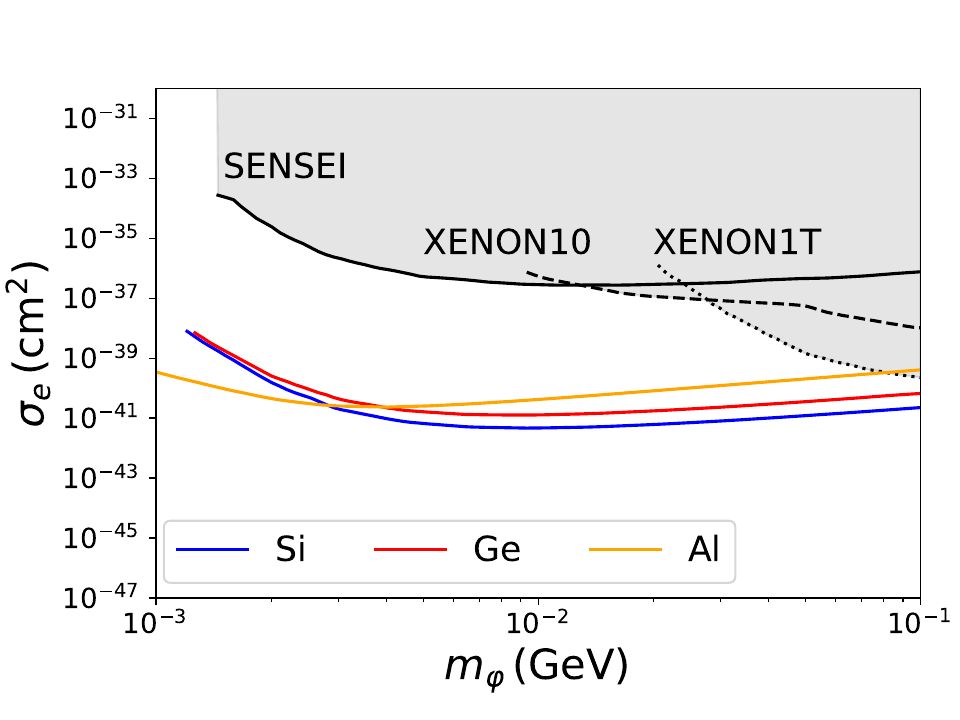}
    \caption{\small \sl Constraints on DM elastic scattering off electrons at 95\% C.L. The shaded regions are excluded by existing direct detection experiments, while the colored lines indicate projected sensitivities of future experiments. See the main text for details.}
    \label{fig: direct detection}
\end{figure}

In the relevant DM mass range, the strongest current bounds come from XENON\,10\,\cite{Essig:2017kqs}, XENON\,1T\,\cite{XENON:2019gfn}, and SENSEI\,\cite{SENSEI:2023zdf}, as shown in Fig.\,\ref{fig: direct detection}. These constraints depend on astrophysical assumptions such as the local DM density and velocity distribution. Here we adopt the benchmark values of Refs.\,\cite{Baxter:2021pqo, Lacroix:2020lhn}, though the limits may be relaxed by up to an order of magnitude under alternative assumptions\,\cite{Baxter:2021pqo, Wu:2019nhd}. Looking ahead, next-generation direct detection experiments are expected to greatly improve sensitivity to DM–electron scattering. Following Ref.\,\cite{Knapen:2021run}, we present projected 95\% C.L. sensitivities for one\,kg$\cdot$year exposures using silicon (Si), germanium (Ge), and aluminum (Al) targets, assuming the same DM distribution. These projections are shown as colored lines in Fig.\,\ref{fig: direct detection}.

% changing colors

%%%%%%%%%%%%%%%%%%%%%%%%%%%
%%%%%%%%%%%%%%%%%%%%%%%%%%%
\subsubsection{Accelerator experiments}
\label{sec: Constraints from Accelerators}

Our model consists of sub-GeV DM, scalar/vector mediators, and the SM particles. The presence of these new states leads to multiple effects: the mediators couple to SM particles via mixing with SM bosons; this mixing modifies the SM interactions involving the corresponding bosons and introduces new interactions between the Higgs boson and the new states. In the following, we summarize the relevant accelerator constraints.  

\begin{figure}[t]
    \centering
    \includegraphics[keepaspectratio, scale=0.41]{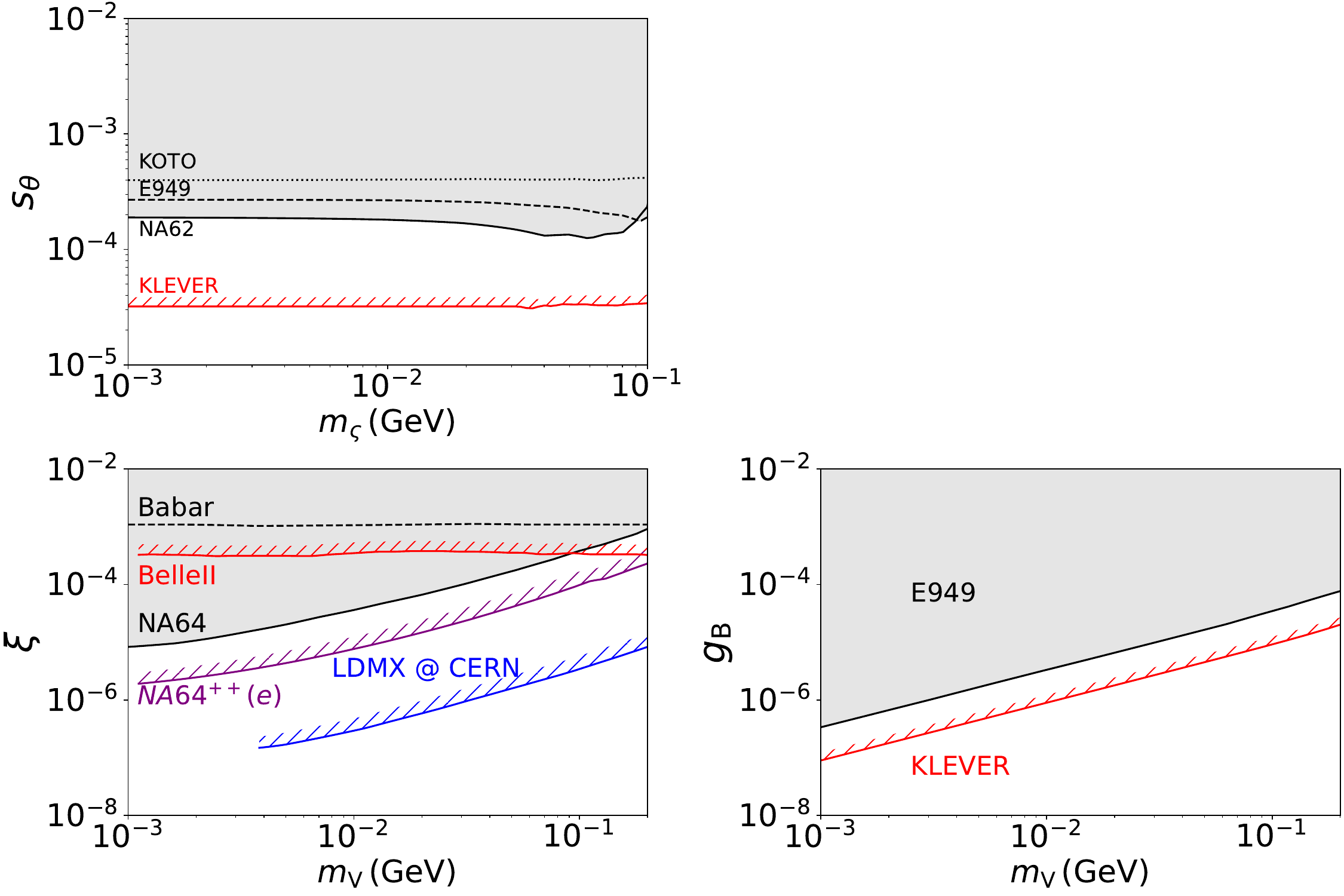}
    \caption{\small \sl Constraints on mediator particles at 90\% C.L.\ from various accelerator experiments. The upper-left panel shows the bounds for the scalar mediator. The bottom panels correspond to the vector mediator: the left panel shows constraints from production via kinetic mixing with neutral gauge bosons, while the right panel shows constraints from baryon-number interactions. See text for details.}
    \label{fig: accelerator constraint}
\end{figure}

\pmb{Scalar mediator production:}
In the mass region of interest, $m_\varsigma \lesssim 100 \, {\rm MeV}$, the scalar mediator is mainly produced in rare kaon decays via the FCNC process $s \to d \varsigma$\,\cite{Krnjaic:2015mbs}. Because its decay width is suppressed by both the small mixing angle $s_\theta$ and the tiny electron Yukawa coupling [see eq.\,(\ref{eq: electrons})], the mediator typically decays outside the detector and thus appears as an invisible particle. The E949\,\cite{BNL-E949:2009dza} and NA62\,\cite{NA62:2021zjw} experiments have searched for $K^\pm \to \pi^\pm + \text{missing}\,(\varsigma)$, while the KOTO experiment\,\cite{KOTO:2020prk} has probed $K^0_L \to \pi^0 + \text{missing}\,(\varsigma)$. The resulting branching-fraction measurements have been recast into bounds on $s_\theta$ following Ref.\,\cite{Krnjaic:2015mbs}, and are summarized in the upper-left panel of Fig.\,\ref{fig: accelerator constraint}. Looking ahead, the KLEVER experiment\,\cite{Beacham:2019nyx} is expected to substantially improve sensitivity through searches for $K^0_L \to \pi^0 + \text{missing}\,(\varsigma)$, as shown by the colored line in the figure.

\pmb{Vector mediator production:}
In the parameter region of interest, the vector mediator decays predominantly into DM and thus appears as an invisible particle. In our model, the vector mediator couples to the SM through two mechanisms: kinetic mixing with neutral gauge bosons and baryon-number interactions. The corresponding constraints are summarized in the bottom panels of Fig.\,\ref{fig: accelerator constraint}. 

For kinetic mixing, the leading current bound comes from the NA64 collaboration\,\cite{Banerjee:2019pds}, which searches for vector mediator production via bremsstrahlung from an electron beam impinging on an active target. In the near future, Belle\,II\,\cite{Belle-II:2018jsg} and NA64$^{++}$\,\cite{gninenko2018addendum} are expected to extend sensitivity by about an order of magnitude, while the proposed LDMX experiment\,\cite{LDMX:2018cma} could probe several orders of magnitude beyond the present reach. 

For baryon-number interactions, anomaly cancellation requires introducing additional heavy fermions (see Sec.\,\ref{sec: scenario}), which in turn induce Wess--Zumino-type interactions between the vector mediator and SM gauge bosons in the low-energy effective theory. These interactions enhance the production of the mediator’s longitudinal mode by a factor $E^2/m_V^2$, where $E$ is the mediator energy\,\cite{Dror:2017nsg, Dror:2017ehi}. The leading bounds are currently set by the E949 collaboration\,\cite{E949:2008btt, E949:2004uaj}, which searched for $K \to \pi V$ decays\,\cite{Ilten:2018crw}. Owing to the longitudinal enhancement, the baryonic coupling $g_{\rm B}$ is strongly constrained. Looking ahead, the KLEVER experiment is expected to improve sensitivity to branching fractions by about an order of magnitude relative to current limits.

%colors?

\pmb{Other accelerator constraints:} 
Modifications of SM interactions induced by mediator mixing have only a negligible impact on constraining our model, since the mixing parameters are already tightly bounded by the accelerator searches discussed above. In contrast, interactions involving the scalar mediator or DM and the Higgs boson can remain sizable even for small mixing angles. These are constrained by LHC measurements of the Higgs invisible decay width, which currently gives ${\rm Br}(h \to \text{inv.}) \leq 0.19$ at 95\% C.L.\,\cite{CMS:2018yfx}.

%%%%%%%%%%%%%%%%%%%%%%%%%%%
%%%%%%%%%%%%%%%%%%%%%%%%%%%
\subsubsection{Indirect detection experiments}
\label{sec: Constraints from Indirect Detection}

Indirect detection experiments aim to observe the products of DM annihilation in space. Depending on the final state, these searches can be broadly classified into two categories: charged particles and neutral particles.

\begin{figure}[t]
    \centering
    \includegraphics[keepaspectratio, scale=0.38]{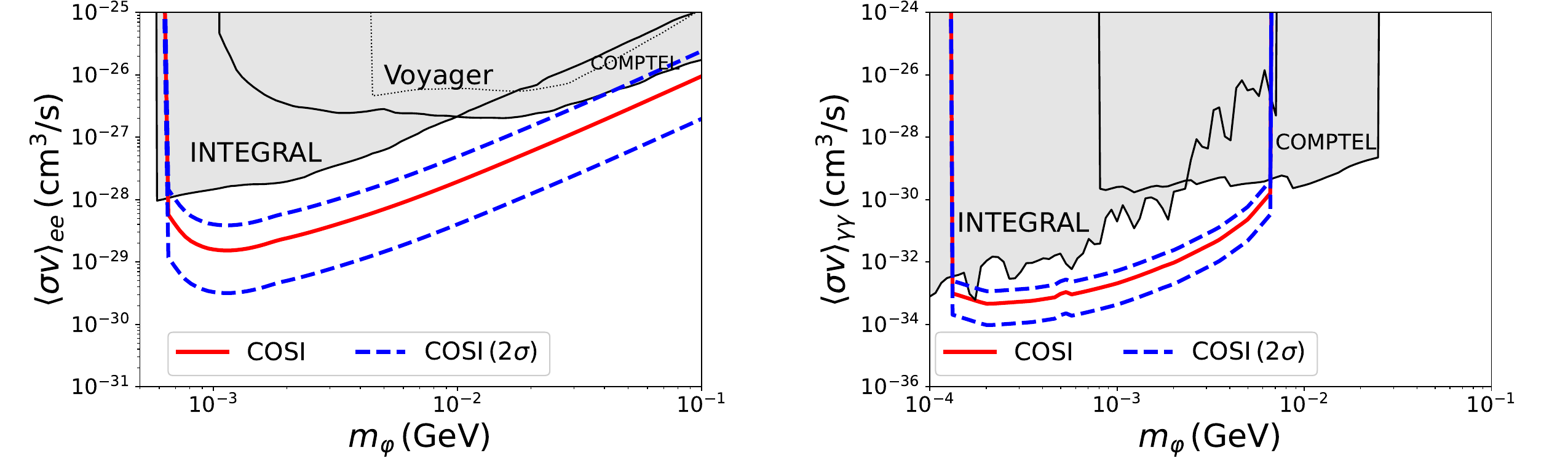}
    \caption{\small \sl 
    Present constraints and projected COSI sensitivities on the DM annihilation cross section into $e^+e^-(\gamma)$ (left) and $\gamma\gamma$ (right). Shaded regions indicate current experimental exclusions. The red solid line shows the COSI sensitivity assuming the median J-factor,\cite{deSalas:2019pee}, while the blue dashed lines illustrate the corresponding $\pm 2\sigma$ uncertainty range.}
    \label{fig: indirect detection}
\end{figure}

\pmb{Charged particle observation:}
For $m_{\varphi} \lesssim 100\,{\rm MeV}$, the annihilation products are electrons and positrons with MeV-scale energies. Experimentally, such particles cannot penetrate the heliosphere because they are deflected by the solar magnetic field\,\cite{Boudaud:2016mos}. The only instrument capable of detecting them is Voyager~I, which has been operating outside the heliosphere since 2012. Theoretically, the predicted signal flux depends on the injected $e^\pm$ spectrum from DM annihilation and its subsequent propagation in the Milky Way, both of which suffer from sizable systematic uncertainties. In particular, variations among propagation models introduce uncertainties spanning several orders of magnitude\,\cite{Donato:2003xg, Boudaud:2016mos, Kappl:2015bqa, Reinert:2017aga}. In this work, we conservatively adopt the results obtained with the propagation model consistent with AMS-02 data and without reacceleration effects, as presented in Ref.\,\cite{DelaTorreLuque:2023olp}. The predicted annihilation signal also depends on the DM density and velocity distribution. Because energetic $e^\pm$ rapidly lose energy through interactions with the interstellar medium, only those produced in the vicinity of the solar system contribute appreciably to the observed flux. We therefore include the uncertainties associated with the local DM density and distribution, adopting the benchmark values from Refs.\,\cite{Baxter:2021pqo, Lacroix:2020lhn}. Rescaling the results of Ref.\,\cite{DelaTorreLuque:2023olp} accordingly, we obtain the 95\% C.L. bound on the annihilation cross section from Voyager~I, shown in the left panel of Fig.\,\ref{fig: indirect detection}.

%colors

\pmb{Neutral particle observation:}
Our DM scenario can also produce MeV $\gamma$-rays via several channels discussed in subsection\,\ref{subsec: Phenomenological quantities}. Experimentally, detecting MeV $\gamma$-rays is challenging due to dominant Compton scattering and complex backgrounds. As a result, the sensitivity in this energy range remains poor compared to other regimes, known as the ``MeV gap''. Thus far, INTEGRAL\,\cite{Winkler:2003nn} and COMPTEL\,\cite{schonfelder1993instrument} have probed this energy window. On the theoretical side, the $\gamma$-ray flux from DM annihilation is estimated by
\begin{align}
    \frac{d{\cal F}_\gamma}{dE_\gamma} \simeq
    \Bigg[
        \frac{\langle \sigma v \rangle}{16 \pi m_\varphi^2} \sum_f\,{\rm Br}\,(\varphi\varphi^* \to f)\,\frac{d N^f_\gamma}{dE_\gamma}
    \Bigg]
    \times
    \Bigg[
        \int_{\Delta \Omega} d\Omega \int_{\rm l.o.s} ds\,\rho_{\rm DM}^2
    \Bigg],
    \label{eq: flux}
\end{align}
where ${\rm Br}\,(\varphi\varphi^* \to f)$ is the branching fraction into the final state $f$, and $dN^f_\gamma/dE_\gamma$ is the photon spectrum per annihilation with energy $E_\gamma$. The second factor on the right-hand side is the J-factor, determined by the DM density profile $\rho_{\rm DM}$. For our analysis, we focus on the GC and adopt the NFW profile\,\cite{Navarro:1995iw, Navarro:1996gj} with the central parameter values given in Table~III of Ref.\,\cite{deSalas:2019pee}. By comparing the predicted flux in eq.\,(\ref{eq: flux}) with the observations\,\cite{Coogan:2021rez, Siegert:2024hmr}, we obtain 95\% C.L. constraints on the annihilation cross section. The left panel of Fig.\,\ref{fig: indirect detection} shows the case of $\varphi\varphi^* \to e^- e^+ (\gamma)$, while the right panel shows $\varphi\varphi^* \to \gamma\gamma$.

Several next-generation Compton telescopes aim to close the ``MeV gap,'' with COSI\,\cite{Tomsick:2021wed, Aramaki:2022zpw, Tomsick:2023aue} being the only currently approved mission. Following the methodology of Refs.\,\cite{Negro:2021urm, Caputo:2022dkz, Tomsick:2023aue}, we estimate the projected COSI sensitivity for two years observations of the GC region within a $10^\circ$ radius, shown in Fig.\,\ref{fig: indirect detection}. The same J-factor as in the COMPTEL/INTEGRAL analysis is used for the red lines. Estimates of the DM density profile\,\cite{deSalas:2019pee, Benito:2020lgu} suggest an order-of-magnitude uncertainty in the J-factor. We therefore vary the parameters within the $2\sigma$ ranges given in Table~III of Ref.\,\cite{deSalas:2019pee} and show the resulting uncertainty bands as blue dashed lines. COSI is expected to improve the sensitivity by more than an order of magnitude compared to current limits.

\begin{figure}[t]
    \centering
    \includegraphics[keepaspectratio, scale=0.65]{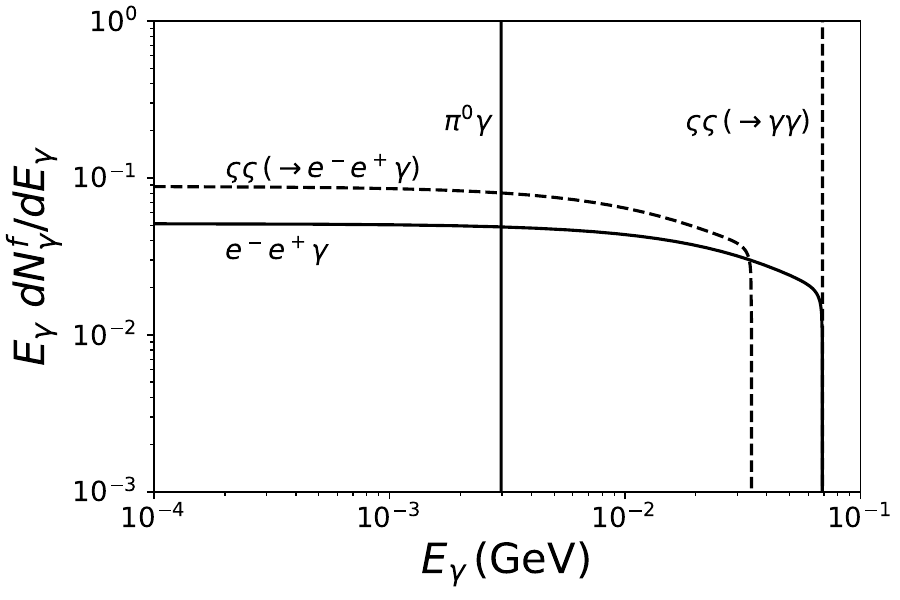}
    \caption{\small \sl Photon spectra from sub-GeV DM annihilation into different channels for $m_\varphi = 69,{\rm MeV}$. Solid lines correspond to annihilation via the vector mediator resonance into $e^+e^-\gamma$ and $\pi^0\gamma$, while dashed lines represent annihilation into scalar mediators that subsequently decay into $e^+e^-\gamma$ and $\gamma\gamma$.}
    \label{fig: dNdE}
\end{figure}

Our model predicts multiple $\gamma$-ray signals, as illustrated in Fig.\,\ref{fig: dNdE}. Annihilation via the vector mediator resonance is strongly enhanced at the GC, since resolving the small-scale crisis requires $v_{\rm R} \sim v_{\rm GC}$ (Sec.,\ref{subsec: self-scattering}). The $e^+e^-\gamma$ channel yields a continuum spectrum with threshold $E_\gamma \simeq m_\varphi$, while the $\pi^0\gamma$ channel produces a line at $E_\gamma \simeq (m_V^2-m_{\pi^0}^2)/(2m_V)$. Twith the subsequent $\pi^0$ decay subdominant compared to the line signal. In addition, DM can annihilate forbiddenly into scalar mediators, which becomes efficient when $v_{\rm th}\lesssim v_{\rm GC}$. Their decays lead either to $e^+e^-\gamma$, generating a continuum with threshold $E_\gamma \simeq m_\varphi/2$, or to $\gamma\gamma$, giving a line at the same energy. Together, these channels produce distinctive spectral features at different energies with sizable fluxes, making this model an attractive target for future MeV $\gamma$-ray astronomy.

%pi0 -> gamma gamma
\subsection{Other constraints}
\label{subsec: Other Constraints}

In addition to the phenomenological bounds discussed above, our scenario is subject to further theoretical and astrophysical constraints. The former include perturbative unitarity and vacuum stability, while the latter arise from supernova (SN) cooling. We summarize these below. 

%%%%%%%%%%%%%%%%%%%%%%%%%%%
%%%%%%%%%%%%%%%%%%%%%%%%%%%
\subsubsection{Perturbative unitarity}
\label{subsec: Perturbative Unitarity}

The conservation of probability in scattering processes is ensured by the unitarity of the $S$-matrix, which in turn requires that the couplings entering the amplitudes not be excessively large\,\cite{Lee:1977eg, Marciano:1989ns}. In the high-energy limit, the dominant contributions to the scattering amplitude come from diagrams involving only dimensionless couplings. Imposing the unitarity condition then yields upper bounds on these couplings, typically of order $\lesssim 4\pi$\,\cite{Arhrib:2011uy, Aoki:2007ah}. By contrast, dimensionful couplings are mainly constrained by scatterings at lower energies and are expected to satisfy approximate bounds of $\lesssim \sqrt{4\pi}$ times the characteristic mass scale of the process\,\cite{Schuessler:2007av, Goodsell:2018tti}. A fully rigorous analysis would lead to somewhat stronger bounds than these heuristic estimates. However, such a comprehensive treatment is technically demanding, even numerically, and the resulting refinements do not significantly affect our conclusions. We therefore adopt the approximate criteria outlined above. 

%%%%%%%%%%%%%%%%%%%%%%%%%%%
%%%%%%%%%%%%%%%%%%%%%%%%%%%
\subsubsection{Vacuum stability}
\label{subsec: Vacuum stability}

The scalar potential consists of the SM Higgs doublet $H$, the dark higgs field $S$, and the DM field $\varphi$, as shown in eq.\,(\ref{eq: lagrangian 1}).  
First, the potential must be bounded from below to ensure that the vacuum energy does not run off to negative infinity, i.e., that the system is well defined.  
This requirement leads to the following constraints on the dimensionless parameters\,\cite{Kannike:2012pe}:
\begin{gather}
    \lambda_H > 0,\quad
    \lambda_S > 0,\quad
    \lambda_\varphi > 0, \quad
    a_{HS} \equiv \lambda_{HS} + \sqrt{\lambda_H \lambda_S} > 0, 
    \nonumber \\
    a_{H \varphi} \equiv \lambda_{H \varphi} + \sqrt{\lambda_H \lambda_\varphi/2} > 0,\quad
    a_{S \varphi} \equiv \lambda_{S \varphi} + \sqrt{\lambda_S \lambda_\varphi/2} > 0,
    \nonumber \\
    \sqrt{\lambda_H \lambda_S \lambda_\varphi /2}
    + \lambda_{HS}\sqrt{\lambda_\varphi/2}
    + \lambda_{H\varphi}\sqrt{\lambda_S}
    + \lambda_{S \varphi}\sqrt{\lambda_H}
    + \sqrt{2 a_{HS} \, a_{H\varphi} \, a_{S \varphi}} > 0.
    \label{eq: vacuum stability BfB}
\end{gather}

Next, the vacuum with expectation values $\langle H \rangle = v_H$, $\langle S \rangle = v_S$, and $\langle \varphi \rangle = 0$ must be, at least, a local minimum.  
This gives rise to the conditions
\begin{align}
    m^2_{\varphi} > 0,
    \quad
    \lambda_H \lambda_S - \lambda_{HS}^2 > 0.
    \label{eq: local minimum cond}
\end{align}

Finally, our vacuum must be sufficiently stable—meaning either absolutely stable or metastable with a lifetime much longer than the age of the Universe.  
Because we impose a $Z_2$ symmetry and the vacuum with $\varphi=0$ is a local minimum, stationary points in the $\varphi$ direction cannot form alternative local minima.  
It therefore suffices to restrict attention to the $(H,S)$ directions, for which the scalar potential reduces to
\begin{align}
    V(h,s)
    = \frac{\lambda_H}{2} \left( h^2 - \frac{v_H^2}{2} \right)^2
    + \lambda_{HS}\left( h^2 - \frac{v_H^2}{2} \right) \left( s^2 - \frac{v_S^2}{2} \right)
    + \frac{\lambda_S}{2} \left( s^2 - \frac{v_S^2}{2} \right)^2,
\end{align}
where we work in the unitarity gauge with $H \equiv (h, 0)^T$ and $S \equiv s$.  
By analyzing the stationary conditions, We find that $V(h,s)$ has stationary points at
\begin{align}
    (h^2,s^2) = (0,0), \, (h_f,0),\, (0, s_f),\, (v_H^2/2,\, v_S^2/2), 
    \quad \text{with}\quad 
    h_f(s_f) = \frac{v_{H(S)}^2}{2} + \frac{\lambda_{HS}}{\lambda_{H(S)}}\frac{v_{S(H)}^2}{2}.
\end{align}
Using eqs.\,(\ref{eq: vacuum stability BfB}) and (\ref{eq: local minimum cond}), we see that 
\begin{align}
    V(0,s_f) = \left(\frac{v_{H}^2}{2}\right)^2\frac{\lambda_H \lambda_S - \lambda_{HS}^2}{2\lambda_S} > 0, 
    \quad \text{and}\quad
    V(0,0) - V(0,s_f) = \frac{\lambda_S}{2}\, s_f^2 > 0.
\end{align}
A completely analogous argument applies to $V(h_f,0)$. Therefore, our vacuum at $(v_H^2/2, v_S^2/2)$ is indeed the global minimum.  

%%%%%%%%%%%%%%%%%%%%%%%%%%%
%%%%%%%%%%%%%%%%%%%%%%%%%%%
\subsubsection{Supernova cooling}
\label{sec: Constraints from Supernova Cooling}

It is well known that the collapse in a SN heats the stellar core, and that neutrino emission is the only SM mechanism capable of cooling the core. In our scenario, however, mediators can also be emitted from the core through their interactions with SM particles. Since they can travel macroscopic distances without transferring energy back to the stellar medium, they provide an additional cooling channel. Because the observed neutrino burst from SN~1987A is in quantitative agreement with SM expectations\,\cite{Burrows:1987zz}, any anomalous energy loss must be sufficiently suppressed\,\cite{Turner:1987by, Burrows:1988ah}. Accordingly, mediator couplings to SM particles are constrained: they must not be so weak that emission is negligible, nor so strong that the mediators become trapped inside the core.

That said, the physics of core-collapse SNe remains highly uncertain\,\cite{Chang:2016ntp, Chang:2018rso, Mahoney:2017jqk, Fischer:2016cyd, Sung:2019xie, Fiorillo:2024upk}. Uncertainties arise from, for example, the microphysics of the proto-neutron-star core, the mechanisms driving shock revival, the structure of the progenitor star (temperature and density profiles, equation of state), the cross sections of various QCD processes with soft radiation, and their environmental modifications. To our knowledge, no study has derived constraints on new particles while systematically incorporating all of these uncertainties. For this reason, we do not include the SN cooling constraint in our analysis, and we leave a robust treatment of this effect to future work.\footnote{Using the SN constraints derived in Refs.\,\cite{Chang:2016ntp, Krnjaic:2015mbs}, we confirmed that although parts of the sub-GeV DM parameter space are excluded, our main conclusions remain unaffected.}

%%%%%%%%%%%%%%%%%%%%%%%%%%%%%%%
%%%%%%%%%%% Results %%%%%%%%%%%
%%%%%%%%%%%%%%%%%%%%%%%%%%%%%%%
\section{The status of self-interacting sub-GeV dark matter}
\label{sec: The Status of Self-Interacting Sub-GeV Dark Matter}

We now turn to the phenomenological implications of our model (Sec.\,\ref{sec: scenario}), together with the conditions and constraints discussed in Sec.\,\ref{sec: conditions and constraints}. Our goal is to identify viable parameter regions that satisfy all requirements, and then to assess the detection prospects in near-future experiments.

%%%%%%%%%%%%%%%%%%%%%%%%%%%
%%%%%%%%%%%%%%%%%%%%%%%%%%%
\subsection{Viable parameter regions}
\label{subsec: Viable Parameter Regions}

\begin{figure}[t]
    \centering
    \includegraphics[keepaspectratio, scale=0.52]{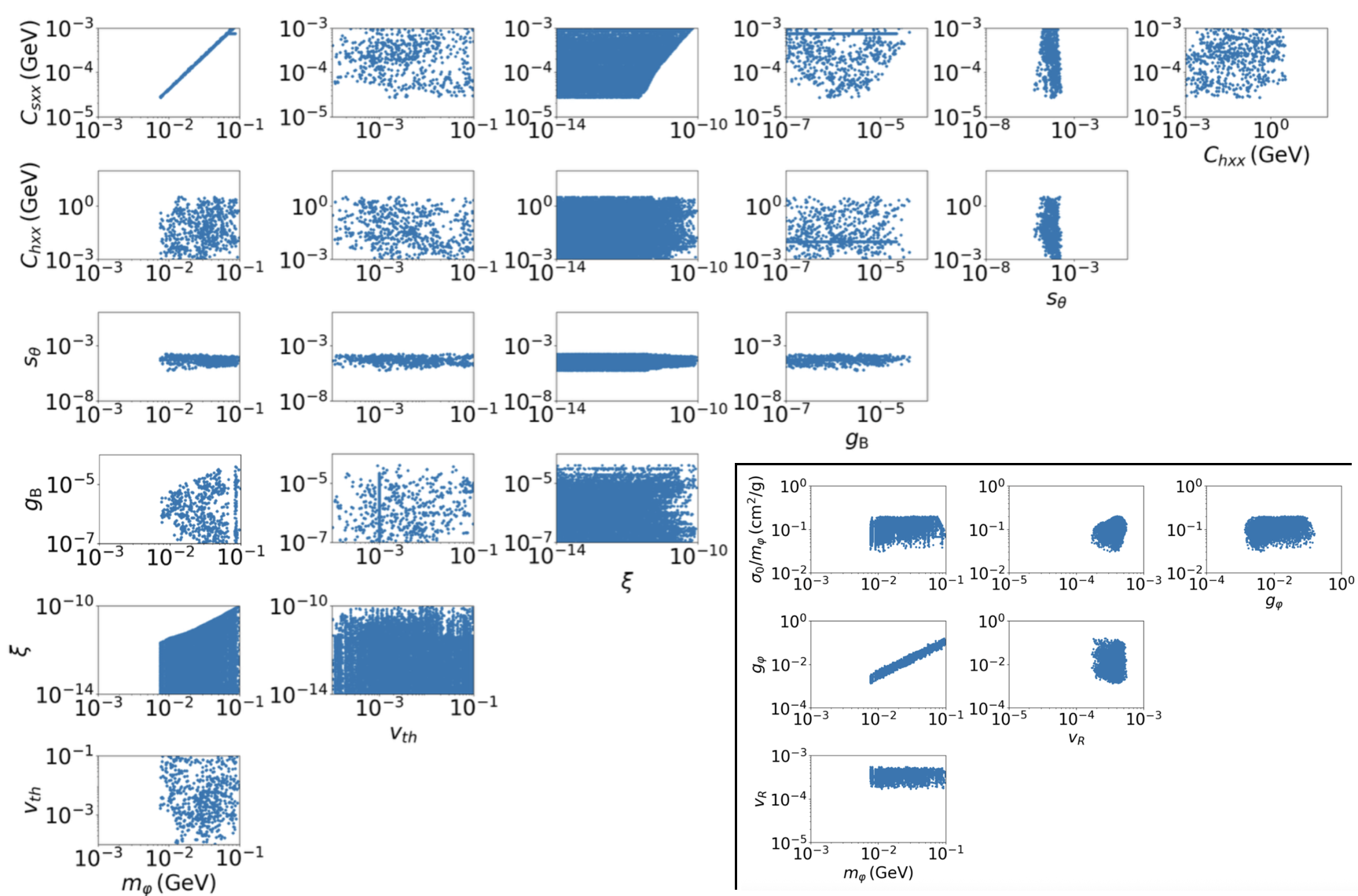}
    \caption{\small \sl Viable parameter region of the sub-GeV DM model that simultaneously reproduces the observed relic density via freeze-out, alleviates small-scale structure problems, and satisfies all cosmological, experimental, and theoretical constraints. See the main text for details.}
    \label{fig: parameter}
\end{figure}

As shown in eq.\,(\ref{eq: lagrangian 1}), the model is originally parametrized by ($m_\varphi$, $v_S$, $\lambda_{S}$, $\xi$, $g_{\rm B}$, $q_\varphi$, $\lambda_{HS}$, $\lambda_{H\varphi}$, $\lambda_{S\varphi}$, $\lambda_{\varphi}$). For phenomenological analyses, however, it is more convenient to work with a set of parameters directly connected to physical observables. We therefore adopt \pmb{$(m_\varphi$, $v_{\rm R}$, $v_{\rm th}$,\, $\xi$, $g_{\rm B}$, $g_\varphi$, $s_{\theta}$, $C_{h\varphi \varphi}$, $C_{\varsigma\varphi \varphi}$, $\sigma_0$)}. We first identify the parameter space favored by the self-scattering requirement and the relic abundance condition. To this end, we performed an MCMC scan of the parameter space using the likelihood functions in eqs.\,(\ref{eq: likelihood self-scattering}) and (\ref{eq: likelihood relic abundance}), implemented with the {\tt emcee} package\,\cite{Foreman-Mackey:2012any}, and determined the regions favored at the 95\% C.L. using the $\Delta\chi^2$ method. We then imposed the constraints discussed in Secs.\,\ref{subsec: Cosmological Constraints}--\ref{subsec: Other Constraints}, removing any parameter sets excluded at the 95\% C.L. by at least one of these constraints.

The resulting viable regions are shown in Fig.\,\ref{fig: parameter}. Because of the large number of parameters, the figure is divided into two panels. The lower-right panel displays the parameters relevant to the self-scattering condition, which requires $g_{\varphi} \propto m_\varphi^{3/2}$, $v_{\rm R} \sim 100\,\mathrm{km/s}$, and $\sigma_0/m_\varphi \sim 0.1\,\mathrm{cm}^2/\mathrm{g}$, as discussed in Sec.\,\ref{subsec: self-scattering}. The remaining parameters are shown in the upper-left panel. The lower bound on the DM mass is about $7.6\,\mathrm{MeV}$, imposed by the CMB constraint discussed in Sec.\,\ref{subsubsec: Constraints from CMB Observation}, while the CMB bound on the annihilation cross section is automatically satisfied unless $v_{\rm th}$ is extremely small ($\lesssim 10^{-5}$). The upper limits on $s_\theta$ and $g_{\rm B}$ arise from mediator searches, while $C_{h\varphi \varphi}$ is constrained by Higgs invisible decay searches at colliders (Sec.\,\ref{sec: Constraints from Accelerators}). Since $v_{\rm R} \sim v_{\rm GC}$, indirect detection experiments provide strong sensitivity (Sec.\,\ref{sec: Constraints from Indirect Detection}): annihilation into $e^-e^+$ imposes stringent bounds on $\xi$, while the $\pi^0\gamma$ channel severely restricts $g_{\rm B}$ when the vector mediator mass is slightly above $m_{\pi^0}$. These bounds are much stronger than those from accelerator searches. In addition, regions with small $m_\varphi$ and $v_{\rm th}$ are excluded by annihilation into mediators that subsequently decay into $e^-e^+$. The kinetic equilibrium condition discussed in Sec.\,\ref{subsec: Relic abundance} imposes a lower bound on $s_\theta$. Taken together, these considerations imply that the $C_{s\varphi\varphi}^2$ term in eq.\,(\ref{eq: Ann forbidden}) dominates the annihilation process at freeze-out, leading to a strong correlation between $m_\varphi$ and $C_{s\varphi\varphi}$ through the relic abundance condition. Furthermore, the local minimum condition of vacuum stability discussed in Sec.\,\ref{subsec: Vacuum stability} requires $m_{\varsigma} > s_\theta m_h$ under these constraints. By contrast, direct detection experiments currently provide no significant bounds, while both perturbative unitarity and bounded-from-below vacuum stability constraints are weak.

%%%%%%%%%%%%%%%%%%%%%%%%%%%
%%%%%%%%%%%%%%%%%%%%%%%%%%%
\subsection{Detection prospects}
\label{subsec: Detection Prospects}

\begin{figure}[t]
    \centering
    \includegraphics[keepaspectratio, scale=0.63]{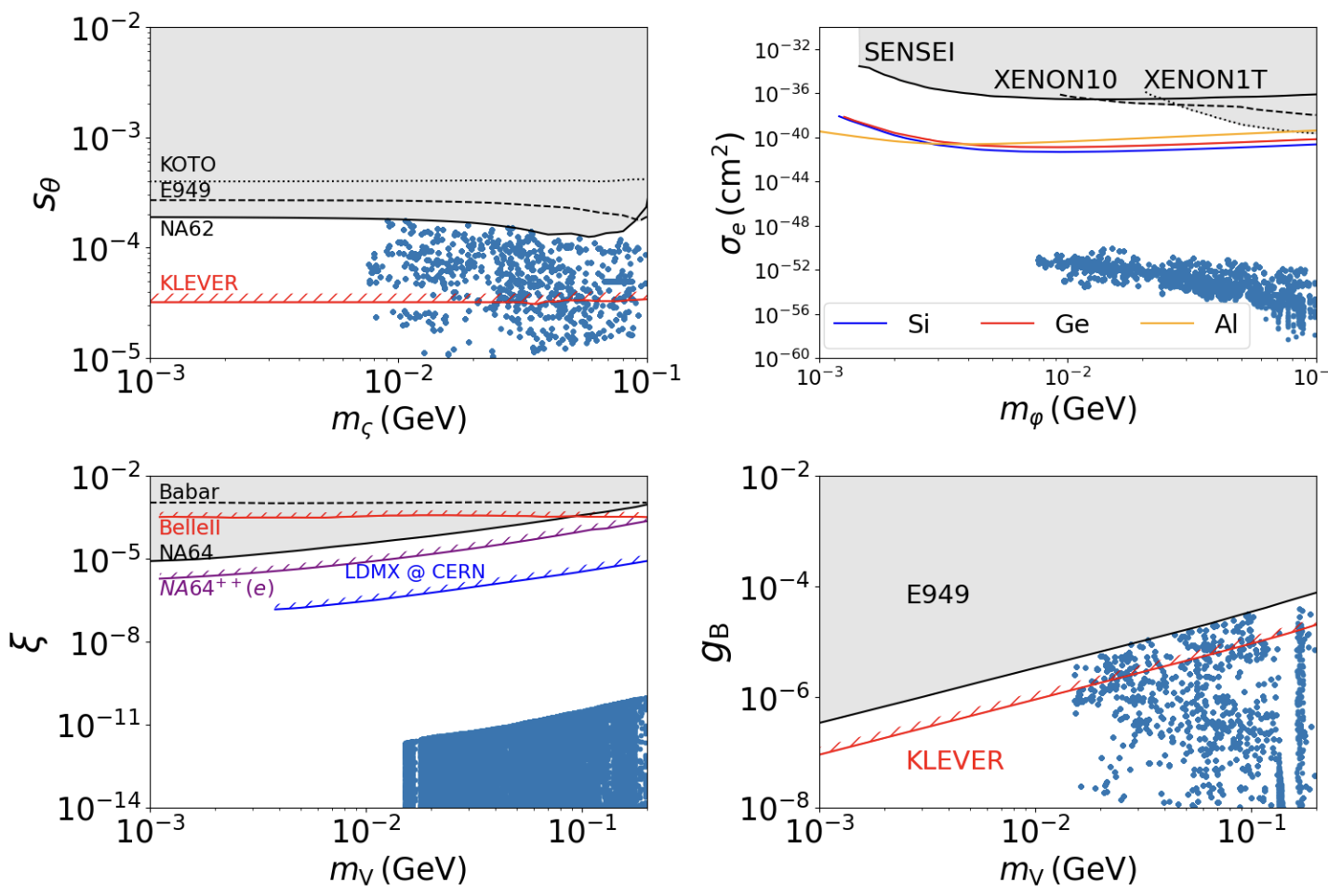}
    \caption{\small \sl Detection prospects for sub-GeV DM at direct detection experiments (upper-left) and mediator searches at accelerator experiments (other three panels). Blue points represent predictions for viable parameter sets, gray shaded regions indicate exclusions from current experiments, and colored solid lines show the projected sensitivities of near-future experiments.}
    \label{fig: detection_DD_col}
\end{figure}

To assess the detection prospects in near-future experiments, we map the viable parameter sets obtained in the previous section onto the observables relevant to the experiments described in Secs.\,\ref{sec: Constraints from Direct Detection}--\ref{sec: Constraints from Indirect Detection}. The prospects for direct detection experiments are shown in the upper-right panel of Fig.\,\ref{fig: detection_DD_col}. The scattering cross section off electrons has two components, as given in eq.\,(\ref{eq: DM-e scattering}); however, the contribution from scalar mediator exchange is suppressed by the tiny electron Yukawa coupling, while that from vector mediator exchange is suppressed by $\xi$, which is already strongly constrained by indirect detection. Consequently, even near-future experiments are unlikely to probe this interaction. The prospects for accelerator experiments are depicted in the other three panels of Fig.\,\ref{fig: detection_DD_col}\footnote{For the vector mediator, only the limiting cases resembling a dark-photon-like scenario ($\xi \gg g_{\rm B}$) or a U(1)$_{\rm B}$-like scenario ($\xi \ll g_{\rm B}$) are shown for illustrative purposes.}. Since $g_{\rm B}$ in the region where the vector mediator mass is slightly above $m_{\pi^0}$ and $\xi$ are already tightly constrained by indirect detection, near-future accelerator searches also have limited reach. Nevertheless, regions with large $s_\theta$ or with large $g_{\rm B}$ in other mass ranges can be probed by the KLEVER experiment.

The prospects for indirect detection through various channels are shown in Fig.\,\ref{fig: detection_DD_col}, where $E_{e^\pm(\gamma)}$ denotes the energy of the emitted electron/positron (photon)\footnote{In actual observations, all channels should be considered simultaneously. In these panels, we present them separately for illustrative purposes, focusing on cases where one channel dominates the others.}. The upper panels correspond to annihilations via the vector mediator resonance into $e^-e^+(\gamma)$ (left) and $\pi^0\gamma$ (right). COSI is expected to probe a significant fraction of the viable parameter space through both continuum and line searches. Notably, if the relic abundance and self-scattering conditions were simultaneously satisfied by a single resonance, an annihilation cross section of $10^{-23}\,\mathrm{cm^3/s}$ at the GC would be required\,\cite{Binder:2022pmf}, which is already excluded by current telescopes. In our framework, however, assigning these roles to two distinct mediators decouples $\xi$ and $g_{\rm B}$ from the relic abundance condition, leaving large regions of parameter space unconstrained. Although the $\pi^0\gamma$ channel is suppressed by $(E_{\gamma}/m_V)^3$ [eqs.\,(\ref{eq: decay pi0gamma}), (\ref{eq: Ann Xsection})] and $g_{\rm B}$ is strongly constrained by accelerator searches, the resonance condition $v_{\rm R} \sim v_{\rm GC}$ can overcome these suppressions and enable a detectable line signal. The lower-left panel shows annihilations into $\varsigma\varsigma$, which subsequently decay into $e^-e^+(\gamma)$, while the lower-right panel shows those into $\gamma\gamma$. We find that COSI will be able to probe a substantial portion of the viable parameter space through continuum searches. Since the annihilation cross section into $\varsigma\varsigma$ is independent of observables from other experiments, COSI can access regions unreachable by alternative probes. While the detection of line-like $\gamma$-ray signals with COSI is not guaranteed, such signals could become observable if the DM profile at the GC is even slightly more cuspy than the NFW profile.

\begin{figure}[t]
    \centering
    \includegraphics[keepaspectratio, scale=0.615]{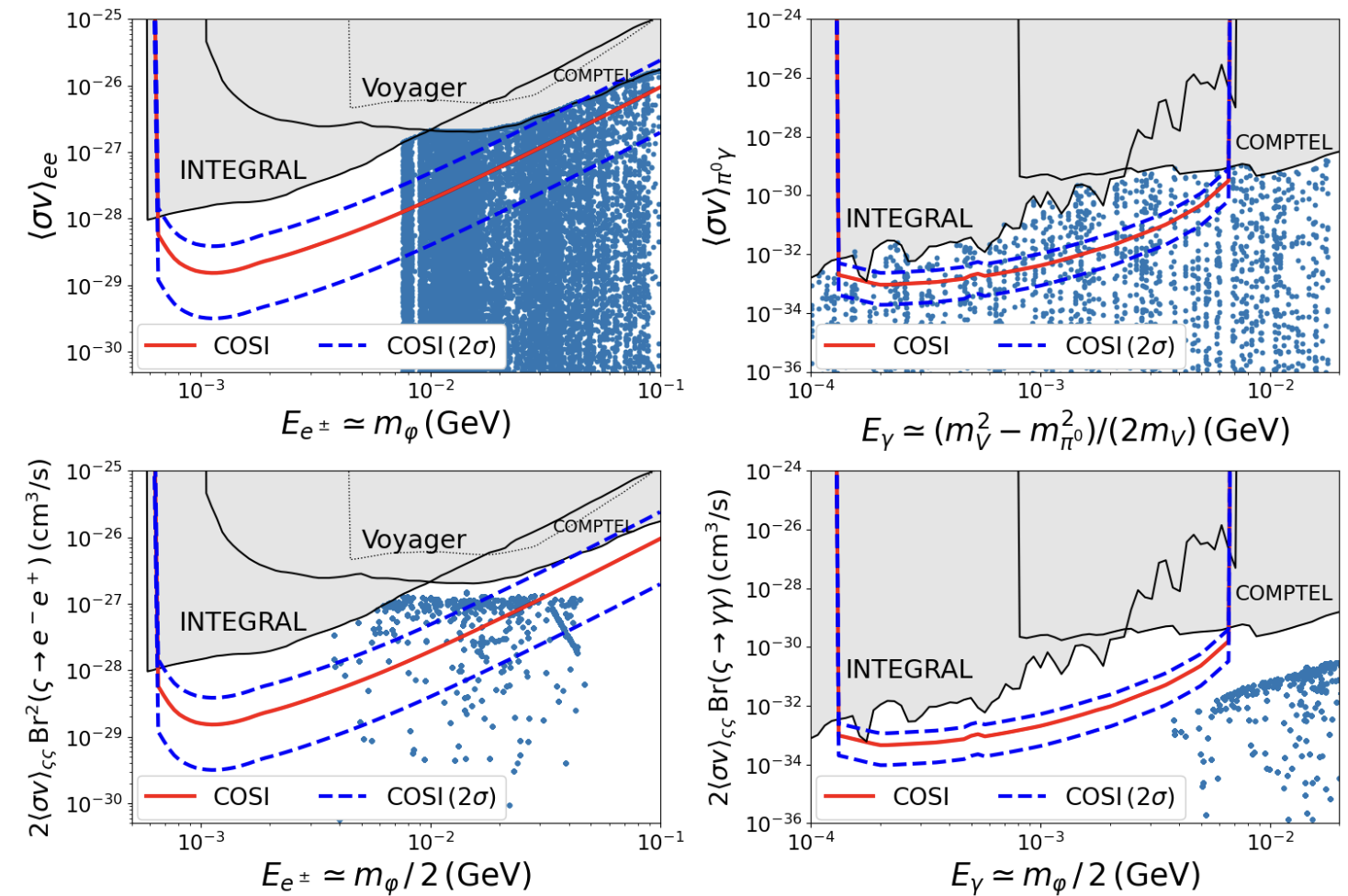}
    \caption{\small \sl Detection prospects for sub-GeV DM at indirect detection experiments. The upper panels correspond to annihilations via the vector mediator resonance into $e^-e^+(\gamma)$ (left) and $\pi^0\gamma$ (right), while the lower panels show forbidden annihilations into scalar mediators, which subsequently decay into $e^-e^+(\gamma)$ (left) and $\gamma\gamma$ (right). Blue points represent predictions for viable parameter sets, and gray shaded regions indicate exclusions from current experiments. The red solid line denotes the COSI sensitivity assuming the median J-factor~\cite{deSalas:2019pee}, while the blue dashed lines indicate the corresponding $\pm 2\sigma$ uncertainty range.}
    \label{fig: detection_ID}
\end{figure}

%xi & gB and line & continuum

%%%%%%%%%%%%%%%%%%%%%%%%%%%%%%%
%%%%%%%%%%% Summary %%%%%%%%%%%
%%%%%%%%%%%%%%%%%%%%%%%%%%%%%%%%%%%%%%%%%%%%%%%%
\section{Summary}
\label{sec: Summary}

Sub-GeV DM has recently attracted increasing attention from both theoretical and experimental perspectives. Among various scenarios, models with $s$-channel resonant enhancement are particularly appealing: they can evade severe constraints from CMB observations while simultaneously addressing small-scale structure issues such as the core–cusp and diversity problems. However, frameworks that attempt to explain both the observed DM abundance via freeze-out and the small-scale problems through the same resonance are strongly constrained by current telescopes, since DM-induced $\gamma$-ray signals from the GC are highly boosted. In this work, we proposed a new framework to overcome these challenges by treating relic abundance and self-scattering requirements independently through two distinct mediators. This decouples the resonance responsible for self-scattering from the annihilation cross section that determines relic abundance, allowing the latter to evade current telescope bounds while still benefiting from resonant enhancement. As an illustrative example, we considered the model with singlet scalar DM that self-scatters via the resonance of a U(1)$_{\rm B}$ vector mediator and undergoes forbidden annihilation into dark Higgs bosons. 

We constructed the general UV-complete Lagrangian of this model and performed a comprehensive scan over all free parameters. We found viable regions that simultaneously reproduce the observed DM relic density via freeze-out, address small-scale structure problems, and satisfy all cosmological, experimental, and theoretical constraints. In particular, CMB observations impose severe limits on DM annihilation, but these are naturally evaded by the resonant and forbidden nature of the annihilation processes. Direct detection experiments provide no significant constraints due to suppressed couplings, while accelerator searches for mediator production and Higgs invisible decays set upper bounds on mediator–SM couplings. Indirect detection experiments impose stringent bounds because of resonance enhancement, whereas theoretical constraints such as perturbative unitarity and vacuum stability turn out to be relatively weak. 

Next, we projected the viable parameter sets onto observables relevant for upcoming experiments. A key result of this work is that a significant fraction of these parameter sets can be probed by the forthcoming MeV $\gamma$-ray telescope, COSI. The model predicts four classes of MeV $\gamma$-ray signatures: DM annihilation via the vector mediator resonance into $e^+e^-$ produces a continuum signal through FSR, while annihilation into $\pi^0\gamma$ produces line signals. In addition, DM forbiddenly annihilates into the scalar mediator, which subsequently decays into $e^-e(\gamma)$ and $\gamma\gamma$. Each channel has a distinctive spectrum and sizable flux, making this model an attractive target for the rapidly advancing field of MeV $\gamma$-ray astronomy. Although we have focused on a specific example, the framework can be straightforwardly generalized to alternative U(1) charge assignments, different spin combinations of DM and mediators, and other annihilation mechanisms, including $p$-wave annihilation.

%%%%%%%%%%%%%%%%%%%%%%%%%%%%%%%%
%%%%%%%%%%% Appendix %%%%%%%%%%%
%%%%%%%%%%%%%%%%%%%%%%%%%%%%%%%%
\appendix
\section{Scalar interactions}
\label{app: scalar interactions}

\begin{align}
	C_{h h h} &= 3\lambda_H v_H c_\theta^3 - 3\lambda_{HS} v_S c_\theta^2 s_\theta - 3\lambda_S v_S s_\theta^3 + 3\lambda_{HS} v_H c_\theta s_\theta^2, \nonumber \\
	C_{\varsigma h h} &= 3\lambda_H v_H c_\theta^2 s_\theta + \lambda_{HS} v_S (c_\theta^3 - 2c_\theta s_\theta^2) + 3\lambda_S v_S c_\theta s_\theta^2 + \lambda_{HS} v_H (s_\theta^3 - 2c_\theta^2 s_\theta), \nonumber \\
	C_{\varsigma \varsigma h} &= 3\lambda_H v_H c_\theta s_\theta^2 + \lambda_{HS} v_S (2c_\theta^2 s_\theta - s_\theta^3) - 3\lambda_S v_S c_\theta^2 s_\theta + \lambda_{HS} v_H (c_\theta^3 - 2c_\theta s_\theta^2), \nonumber \\
	C_{\varsigma \varsigma \varsigma} &= 3\lambda_H v_H s_\theta^3 + 3\lambda_{HS} v_S c_\theta s_\theta^2 + 3\lambda_S v_S c_\theta^3 + 3\lambda_{HS} v_H c_\theta^2 s_\theta, \nonumber \\
	C_{h h h h} &= 3\lambda_H c_\theta^4 + 6\lambda_{HS} c_\theta^2 s_\theta^2 + 3\lambda_S s_\theta^4, \nonumber \\
	C_{\varsigma h h h} &= 3\lambda_H c_\theta^3 s_\theta - 3\lambda_{HS} (c_\theta^3 s_\theta - c_\theta s_\theta^3) - 3\lambda_S c_\theta s_\theta^3, \nonumber \\
	C_{\varsigma \varsigma h h} &= 3\lambda_H c_\theta^2 s_\theta^2 +\lambda_{HS} (c_\theta^4 - 4c_\theta^2 s_\theta^2 + s_\theta^4) + 3\lambda_S c_\theta^2 s_\theta^2, \nonumber \\
	C_{\varsigma \varsigma \varsigma h} &= 3\lambda_H c_\theta s_\theta^3 + 3\lambda_{HS} (c_\theta^3 s_\theta - c_\theta s_\theta^3) - 3\lambda_S c_\theta^3 s_\theta, \nonumber \\
	C_{\varsigma \varsigma \varsigma \varsigma} &= 3\lambda_H s_\theta^4 + 6\lambda_{HS} c_\theta^2 s_\theta^2 + 3\lambda_S c_\theta^4, \\
    \nonumber \\
    C_{h \varphi \varphi} &= \lambda_{H \varphi} v_H c_\theta - \lambda_{S \varphi} v_S s_\theta, \nonumber \\
    C_{\varsigma \varphi \varphi} &= \lambda_{H \varphi} v_H s_\theta +  \lambda_{S \varphi} v_S c_\theta, \nonumber \\
    C_{h h \varphi \varphi} &= \lambda_{H \varphi} c_\theta^2 + \lambda_{S \varphi} s_\theta^2, \nonumber \\
    C_{h \varsigma \varphi \varphi} &= \lambda_{H \varphi} c_\theta s_\theta - \lambda_{S \varphi} s_\theta c_\theta, \nonumber \\
    C_{\varsigma \varsigma \varphi \varphi} &= \lambda_{H \varphi} s_\theta^2 + \lambda_{S \varphi} c_\theta^2.
    \label{eq : couplings}
\end{align}

%%%%%%%%%%%%%%%%%%%%%%%%%%%%%%%%%%%%%%%
%%%%%%%%%%% Acknowledgments %%%%%%%%%%%
%%%%%%%%%%%%%%%%%%%%%%%%%%%%%%%%%%%%%%%
\section*{Acknowledgments}

The author thanks Shigeki Matsumoto for reviewing the draft and acknowledges support from the JSPS Overseas Research Fellowships and the U.S. Department of Energy Award Number DE-SC0009937.

%%%%%%%%%%%%%%%%%%%%%%%%%%%%%%%%%%
%%%%%%%%%%% References %%%%%%%%%%%
%%%%%%%%%%%%%%%%%%%%%%%%%%%%%%%%%%
\bibliographystyle{unsrt}
\bibliography{refs}

\end{document}